\documentclass{vgtc}                          

\ifpdf
  \pdfoutput=1\relax                   
  \usepackage{graphicx}                
  \DeclareGraphicsExtensions{.pdf,.png,.jpg,.jpeg} 
\else
  \usepackage{graphicx}                
  \DeclareGraphicsExtensions{.eps}     
\fi%

\graphicspath{{figures/}{pictures/}{images/}{./}} 

\usepackage{microtype}                 
\PassOptionsToPackage{warn}{textcomp}  
\usepackage{textcomp}                  
\usepackage{mathptmx}                  
\usepackage{times}                     
\usepackage{cite}                      
\usepackage{tabu}                      
\usepackage{booktabs}                  
\usepackage{subfig}
\usepackage[export]{adjustbox}
\usepackage{amsmath,amssymb}
\usepackage{mathtools}

\DeclarePairedDelimiter\norm{\lVert}{\rVert}%

\newcommand{\Vor}{\textrm{Vor}}

\usepackage{xcolor}

\newcommand{\eat}[1]{}

\newcommand{\mb}[1]{\mathbf{#1}}

\onlineid{1718}

\vgtccategory{Research}

\vgtcinsertpkg

\title{Visualization of Unsteady Flow Using Heat Kernel Signatures}

\author{Kairong Jiang\thanks{e-mail: jiangkairong@email.arizona.edu}\\ %
        \scriptsize University of Arizona %
\and Matthew Berger\thanks{e-mail: matthew.berger@vanderbilt.edu}\\ %
      \scriptsize Vanderbilt University
\and Joshua A. Levine\thanks{e-mail: josh@email.arizona.edu}\\ %
      \scriptsize University of Arizona }

\teaser{
  \centering
  \includegraphics[width=0.75\linewidth]{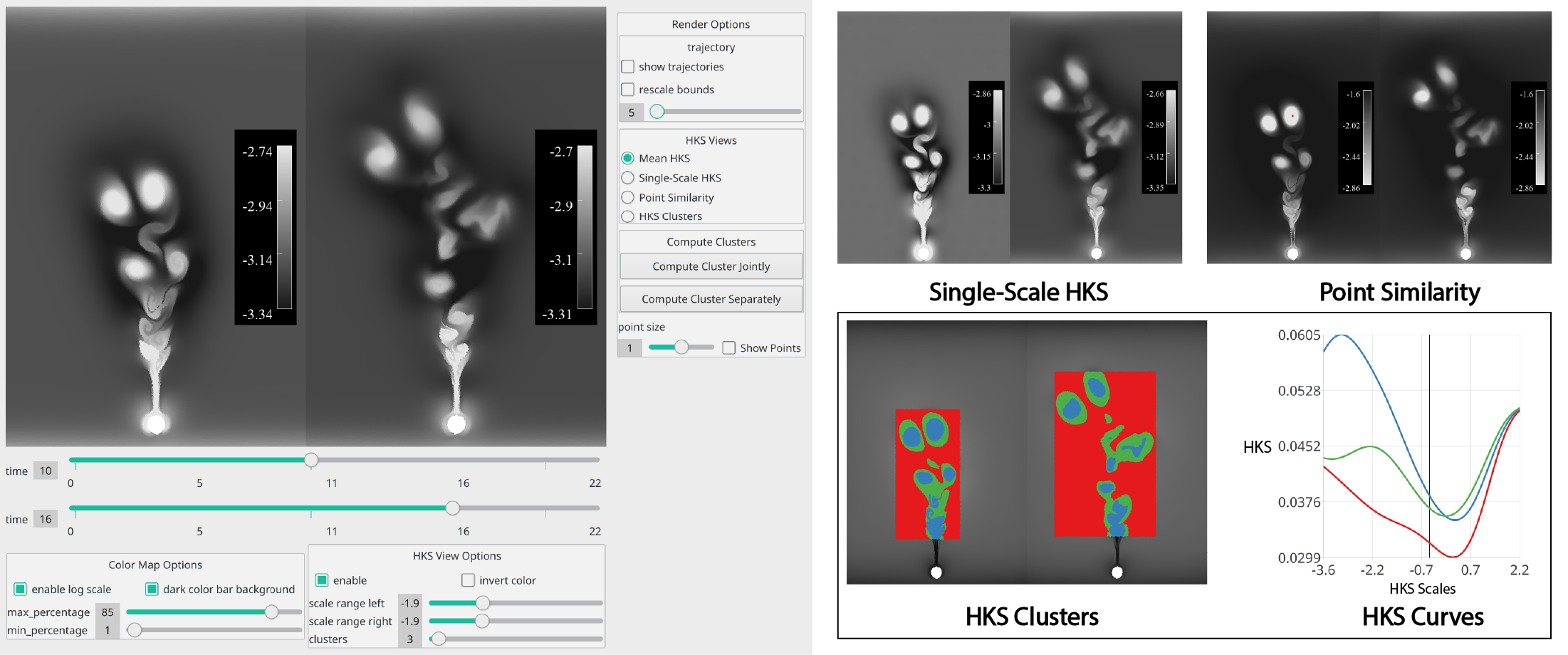}
  \caption{We visualize the Heat Kernel Signature (HKS) of flow pathlines. Left: an overview of our interface, showing Mean HKS View of two timesteps from a heated cylinder dataset. Right: the same two timesteps are visualized using the Single-Scale HKS, Point Similarity, and HKS Cluster Views while displaying HKS Curves for the centroids of the clusters. \label{fig:teaser}}
}

\abstract{
We introduce a new technique to visualize complex flowing phenomena by using concepts from shape analysis.  
Our approach uses techniques that examine the intrinsic geometry of manifolds through their heat kernel, to obtain representations of such manifolds that are isometry-invariant and multi-scale. 
These representations permit us to compute heat kernel signatures of each point on that manifold, and we can use these signatures as features for classification and segmentation that identify points that have similar structural properties.
Our approach adapts heat kernel signatures to unsteady flows by formulating a notion of shape where pathlines are observations of a manifold living in a high-dimensional space.
We use this space to compute and visualize heat kernel signatures associated with each pathline.
Besides being able to capture the structural features of a pathline, heat kernel signatures allow the comparison of pathlines from different flow datasets through a shape matching pipeline.  
We demonstrate the analytic power of heat kernel signatures by comparing both (1) different timesteps from the same unsteady flow as well as (2) flow datasets taken from ensemble simulations with varying simulation parameters.  
Our analysis only requires the pathlines themselves, and thus it does not utilize the underlying vector field directly.  
We make minimal assumptions on the pathlines: while we assume they are sampled from a continuous, unsteady flow, our computations can tolerate pathlines that have varying density and potential unknown boundaries. 
We evaluate our approach through visualizations of a variety of two-dimensional unsteady flows. 
} 


\CCScatlist{
  \CCScatTwelve{Human-centered computing}{Visu\-al\-iza\-tion}{Visu\-al\-iza\-tion application domains}{Scientific visualization};
}

\begin{document}


\firstsection{Introduction}

\maketitle





In this paper we are motivated by the following problem: given an unsteady flow dataset with pathlines, how do we measure the structural similarity between pathlines in terms of their temporal and spatial behavior?
Pathlines offer a rich descriptor of individual datasets, so by analyzing the structural properties of pathlines one may be able to distinguish the underlying flow behaviors and conduct important analysis tasks. 

We are inspired by a similar challenge in the shape analysis community: given a geometric shape (with its bounding surface as a discrete mesh or point set), one may want to measure similarities or find correspondences between shapes. 
An attractive approach is to analyze the heat diffusion process on the shape, which characterizes the \emph{neighborhood} of a point by recording the dissipation of heat from that point to the rest of the shape over time. 
This process is intrinsic to the shape geometry, and provides a natural notion of scale through its time-progressive nature.  For short spans of time, heat diffuses only in the most local of neighborhoods and captures local curvature variations, while for large spans of time two points will have similar diffusion if they are coarsely similar.

To model this process, \emph{heat kernels} capture all the information of heat diffusion on a shape, and provide an isometry-invariant and multi-scale descriptor of points on a shape~\cite{sun2009concise}. Notably, heat kernels have a scale parameter that governs how much heat may diffuse on the shape between any pair of points.
Built upon the heat kernel, the \emph{heat kernel signature} (HKS), quantifies the amount of heat diffusion at a single point across many scales. It inherits the intrinsic properties of heat kernel, while being concise and stable under perturbations of the shape~\cite{sun2009concise}.  
These features have made the HKS popular for applications in shape comparison,
segmentation~\cite{dey2010persistent} and
correspondence~\cite{ovsjanikov2010one}. 

We use the HKS to characterize the structural properties of pathlines in a multi-scale manner that also enables comparison and classification between pathlines of a single flow or even different flows. Concretely, we define a notion of shape with regard to the pathlines by taking them as points from some manifold that lives in the high-dimensional space of the pathlines.  Then, as others do, we compute the discrete Laplace-Beltrami operator (LBO) on the manifold approximated by the given pathlines, and compute the HKS for each pathline.

The HKS of the pathlines has several attractive properties that translate to flow analysis. Firstly, since it is isometry-invariant and robust to small perturbations from isometry, we can compare flow behaviors that are structurally similar, yet lie in different spatial regions or move along different trajectories. Secondly, the HKS is not merely a high-dimensional feature, but also may be plotted as a smooth 1D curve that the user can interpret and compare for different pathlines.  This comparison can be in aggregate by asking how different HKS curves are at all scales, or it can be reduced to comparisons at specific ranges of scales.  Finally, the HKS is not application-specific, but rather it is a mathematical description of the abstract manifold on which the pathlines lie, as opposed to a direct expression of physical quantities such as divergence or vorticity.  We see this as an advantage for a complementary exploration of flow data that captures variation in the space of pathlines and can be used to visualize behaviors that are similar or dissimilar.  

Another advantage of the HKS is that it is multi-scale, meaning that it characterizes a pathline with respect to different spatial scales simultaneously. Other techniques that compute a scalar property for the pathlines typically compute measures that consider infinitesimally small changes in space and time for vector field data, for instance, when extracting vortices~\cite{jiang2004detection} by using vorticity~\cite{jeong1995identification} or the eigenvector method~\cite{sujudi1995identification}. Other methods such as the finite-time Lyapunov exponent (FTLE) compute measures of separation/attraction of pathlines at instantaneous change in space, and a user-provided time range~\cite{shadden2005definition}. These existing techniques share the limitation of \emph{fixed spatial scale}: they compute quantities that assume a fixed, instantaneous rate of spatial change, e.g.~spatial derivatives. This limits their discriminating power for comparing pathlines. For instance, consider two pathlines that lie in different regions of mainly laminar flow. One of the pathlines may be closer to a region where the flow is transitioning from laminar to turbulent. Approaches that assume a fixed spatial scale in their analysis may fail to distinguish these types of configurations, instead reporting the pathlines as being similar because, locally, they appear laminar. We would like to have a more precise comparison between different flows, e.g.~the ability to discover the size and shape of regions of transition between structures where vortex shedding is occurring. 

We summarize our contributions in the following:

\begin{itemize}
	\item We use the heat kernel as a method for studying the intrinsic geometry of unsteady flows.
	\item We provide techniques to compute the Laplace-Beltrami operator of the manifold which the pathlines approximate, and evaluate both computational considerations as well as technical issues with boundaries and the spatial distribution of pathlines.
	\item We extract the HKS to visualize flows, enabling us to identify interesting, multi-scale behaviors as well as to jointly cluster and compare different datasets.
\end{itemize}

\section{Related Work}




A popular approach to visualizing flow integrates massless particles and produces a geometric set of features that are the paths these particles traverse~\cite{McLoughlinLPPC10}.  Our work falls into this class of techniques as well, so we briefly review key concepts.

\paragraph{Visualizing the structure of pathlines}

In the case of a steady flow dataset, these paths are typically integrated maximally, producing \emph{streamlines}, whereas in the unsteady case one considers \emph{pathlines}.  Streamlines are comparatively easier to visualize, as they do not cross.  As a result, one can use a static representation to describe the space of streamlines.  
Helman and Hesselink's \emph{vector field skeleton}~\cite{HelmanH89}, together with \emph{closed orbits}~\cite{wischgoll:closedorbits}, captures this concept, as it provides a structural description of all possible streamlines. 
Elements of the skeleton represent bounding paths that, in the limit, encode separations between bundles of streamlines.  Alternatively, Morse decompositions describe the full segmentation~\cite{ChenMLZ08, ChenDSLZ12} into groups of streamlines that move similarly by partitioning the domain itself.  In higher dimensions, these structures are much more complex~\cite{GlobusLL91}.  For a full survey of the use of topological analysis in flow visualization, see Garth and Tricoche~\cite{GarthT05}, Scheuermann and Tricoche~\cite{ScheuermannCH04} and Laramee et al.~\cite{LHZP07}.

Switching to the unsteady case has pushed the limits of studying such geometric flow features~\cite{PobitzerPFSKTMH11}, leading to successes on tracking topological features~\cite{GarthTS04} as well as defining pathline~\cite{TheiselWHS04,shi2006path} and streakline~\cite{UffingerSE13} topology.  Recently, an alternative characterization of unsteady flow topology has focused on studying separation through the \emph{Finite-Time Lyapunov Exponent} (\emph{FTLE})~\cite{BarakatGT12, FuchsKSWSHP10, haller2001distinguished}. 
The ridges of the FTLE approximate \emph{Lagrangian Coherent Structures} (\emph{LCS})~\cite{shadden2005definition}, a time-varying counterpart to the arcs of the vector field skeleton.
Intuitively, the FTLE considers the flow map, a function that identifies where a particle will travel in a fixed time span, $\tau$, and evaluates its spatial derivatives.
We have the same constraints of a fixed time span, but our approach enables a multi-scale interpretation of neighborhoods instead of just the derivative.

\paragraph{Extracting temporal features from unsteady flow}
While a concise visual summary, flow topology is still considered expensive to compute and requires assumptions on numerical behaviors, sampling, and integration.  Furthermore, for time-varying 3D flow fields, its definition is incomplete when it comes to describing the three-dimensional equivalent of closed orbits.   
Subsequently, recent research has considered more general structures in vector fields by analyzing sets of flow paths, but most techniques still only address steady flow.
R{\"o}ssl and Theisel measure similarity of streamlines by Hausdorff distance, embed them into 2D/3D via multidimensional scaling, and perform spectral clustering~\cite{RosslT12}.  They show connections to the topological decomposition under infinite integration of the underlying vector field.  Others consider hierarchical grouping of streamlines via agglomerative clustering~\cite{YuWSC12}, using cluster exemplars to summarize sets of streamlines~\cite{OeltzeLKJTP14}, clustering in parallel~\cite{wei2011parallel}, deep learning for clustering~\cite{han2018flownet}, and Gaussian mixture models to visualize ensembles of streamline sets~\cite{ferstl2016streamline}.  

Different approaches have addressed other types of features and techniques to compare streamlines. Lu et al.~propose to first segment individual streamlines, using curvature, curl, and torsion as features for each segment, and compare streamline segments via dynamic time warping~\cite{LuCLSW13}.  Tao et al.~develop a vocabulary-based approach to measure partial similarities~\cite{TaoWS14}, while Chaudhuri et al.~use space-filling curve complexity measures~\cite{ChaudhuriLSW14}.  Such approaches can infer both local and global patterns for users~\cite{WangESW14}.  
Interactive techniques based on graph metaphors~\cite{MaWSJ14} or example-based sketching~\cite{Li2015,LuCLSW13} help design such features. 

The previously mentioned works motivate a feature-based analysis of flow, but unlike us they do not employ feature spaces that respect the time dimension.  Relatively few works do consider temporal features.
Wei et al. consider analysis in the context of time-varying combustion simulations~\cite{wei2012visual}.  Sauer et al. employ features defined across Eulerian-Lagrangian frames~\cite{sauer2014trajectory,sauer2017spatio} and Zhang et al.~accumulate Lagrangian characteristics~\cite{ZhangLTSC16}.  
Berenjkoub et al. consider the pairwise relationships between attributes in unsteady flow that respect the temporal component~\cite{berenjkoub2019visual}.  Hong et al.~apply latent Dirichlet allocation to model pathlines as mixtures of topics, where topics are defined as mixtures of time-indexed user-provided features~\cite{HongLGSYL14}.  Finally, Guo et al.~project pathlines of different time intervals into a common 2D space~\cite{guo2014scalable}.  While many of these share related goals to the current work, they mainly take an additive approach to designing the feature spaces (i.e.~position + velocity values) as opposed to a modeling process that learns the desirable properties of the feature space (e.g.~intrinsic representations).

\paragraph{Understanding Temporal Scale}

A key component of our approach is integrating a notion of multi-scale temporal features that describe pathlines.  
 This is related to scale-space approaches~\cite{lindeberg2013scale} which analyze field-based data in multiple scales, typically to denoise or find optimal spatial scales for filtering. Flow analysis has employed scale-space techniques in various contexts, such as vortex tracking~\cite{bauer2002vortex} and detection of FTLE ridges~\cite{fuchs2012scale}.  However, the construction of multi-scale distances requires different mathematical tools compared to traditional scale-space methods on fields.  Furthermore, it is nontrivial to extend these techniques to pathline data, as they utilize a vector field for their respective scale-space approaches.



Heat kernels are intimately related to diffusion, and a related set of research has focused on diffusion-type operators on flow data, and have been employed to extract a notion of coherent sets~\cite{Froyland15}, defined as spatial regions that are robust to small perturbations applied to the flow map used in FTLE.  Related works model the perturbation as a diffusion process by constructing different types of diffusion objects such as the Laplace operator~\cite{froyland2016dynamic},
the heat kernel~\cite{karrasch2016geometric}, and space-time diffusion maps~\cite{banisch2017understanding}, for downstream use in clustering coherent sets~\cite{hadjighasem2016spectral}.  Diffusion maps have also been used to define separation and similarity for pathlines, going beyond just clustering as a visualization primitive~\cite{berger2017visualizing}. Our work assumes a similar model, e.g. a set of pathlines modeled as a manifold embedded in a high-dimensional space. However, instead of using the manifold to group tightly-bundled pathlines~\cite{banisch2017understanding}, we construct shape signatures that characterize the manifold's intrinsic geometry, which enables a very different type of analysis.

\section{Background}

\subsection{Objectives}
Our primary objective in this paper is to adapt techniques that have been developed by the shape analysis community to the problem of unsteady flow, so that we may construct a representation of flow that satisfies the following goals:
\begin{itemize}
	\item (G1) \textbf{Multi-scale}: we would like our approach to capture features at different levels of spatial scale, in a manner where the user can interpret features with respect to scale, and subsequently, visually explore flow data through scale selection.
  \item (G2) \textbf{General Purpose}: instead of focusing on a prescribed set of features (e.g. vorticity or shear) that describe physical properties, we aim to model a generic feature that captures behavioral variation. 
	\item (G3) \textbf{Commensurable}: the representation of flow data should permit a direct comparison of derived features both within an individual dataset, as well as across datasets.
\end{itemize}
To this end, we use the \emph{heat kernel} to analyze the \emph{intrinsic geometry} of unsteady flows. 
The heat kernel of a surface captures how much heat flows between a pair of points, parameterized by a scale parameter that governs the spatial scale at which heat may diffuse. The heat kernel is useful for shape analysis as it is invariant to isometries, robust to small geometric perturbations~\cite{sun2009concise}, and can be used to capture surface features at varying scales. 

For example, suppose we were given a surface representing a hand, and we wanted to understand the similarity of the tips of two fingers. At sufficiently small scales, the flow of heat from the two fingertips to the rest of the hand would be limited to their respective fingers. Therefore, these fingertips would be considered similar under these scales. For sufficiently large scales, however, heat would flow to the remaining geometry of the hand, indicating that these fingertips are, indeed, different (for example, fingers of different lengths would have different heat diffusions).

We analyze unsteady flow under the lens of the heat kernel. To this end, we represent an unsteady flow dataset as a collection of its pathlines, and develop a notion of shape by considering this set of pathlines as approximating an underlying  manifold. We then construct the heat kernel with respect to this manifold, and extract features of this heat kernel, namely the heat kernel signature (HKS)~\cite{sun2009concise}, to study features in unsteady flow.  Since we start with a set of pathlines that are unstructured, we utilize a discrete Laplace-Beltrami operator (LBO) that we approximate using the method of Liu et al.~\cite{6264046}.

The HKS has a built-in notion of scale as it is derived from the heat kernel, enabling us to study flow features under different scales on the manifold (G1).
The HKS is a general shape signature of the manifold formed by the set of pathlines, and thus allows us to capture aspects of multiple types of common flow features (G2). Last, the HKS is a function of the intrinsic geometry of the manifold, and thus we may extract the HKS from separate datasets and directly compare them in terms of the HKS (G3).

\subsection{Mathematical Preliminaries}
\label{subsec:prelims}

We assume that we are provided a set of $n$ pathlines representing an unsteady flow starting at time $t_0$ integrated over a temporal interval $[t_0,t_0+\tau]$. Each pathline $\mb{p}$ is described by concatenating its positions at a set of $m$ timesteps sampled from $[t_0,t_0+\tau]$. i.e. we have $m$ timesteps ${t_0, t_1, \ldots, t_{m-1}}$, where $t_{m-1} = t_0+\tau$, and we have $\mb{p} = (\mb{x}_{t_0}, \mb{x}_{t_1}, \ldots, \mb{x}_{t_{m-1}})$, where $\mb{x}_{t_i}$ denotes the position of pathline $\mb{p}$ at timestep $t_i$.
We assume that every pathline has the same time discretization, i.e. the same set of $m$ timesteps.  

We view the pathlines as points on a manifold $P$ living in $(d \times m)$-dimensions. That is, we have a point set $\{\mb{p} \in P\}$ approximating the manifold $P$. The typical method for computing the heat kernel is to consider its relationship to the Laplace-Beltrami operator (LBO) through the heat equation:
\begin{equation}
	\Delta_P u(\mb{p},s) = -\frac{\partial u(\mb{p},s)}{\partial s},
\end{equation}
where $u(\mb{p},s)$ is the heat equation which describes how heat diffuses over time from pathline $\mb{p}$ at a certain scale $s$, and $\Delta_P$ is the LBO as constructed on manifold $P$. The LBO is defined as the divergence of the gradient. That is, $\Delta_P f = \nabla_P \cdot \nabla_P f$.

This equation governs the process of heat diffusion on the manifold for a prescribed scale $s$, which determines the spatial extent to which heat may diffuse -- heat diffuses less for smaller scales, and more for larger scales. The solution of the heat equation $u$, may be expressed in terms of some initial function $f_0 : P \rightarrow \mathbb{R}$ that serves as the boundary condition, and the heat operator $H_s$:
\begin{equation}
	H_s f_0(\mb{p}) = \int_{P} k_s(\mb{p},\mb{q}) f(\mb{q}) d \mb{q},
\end{equation}
where $k$ is the \emph{heat kernel} on the manifold $P$. It is a function $k : \mathbb{R}^{+} \times P \times P \rightarrow \mathbb{R}$, that describes the amount of heat that has been transferred from
one pathline $\mb{p}$ to another $\mb{q}$ under a given spatial scale $s$. In particular, it may be expressed in terms of the eigenvalues $\lambda_i$ and eigenvectors $\phi_i$ of $\Delta_P$~\cite{hsu2002stochastic}:
\begin{equation}\label{eq:heat_kernel}
	k_s(\mb{p},\mb{q}) = \sum_{i=0}^{m} e^{-\lambda_i s} \phi_i(\mb{p}) \phi_i(\mb{q}).
\end{equation}

\paragraph{From Heat Kernels to HKS}


\begin{figure}[!t]
  \centering
  \subfloat[\label{fig:heat_kernel_small}]{
    \includegraphics[width=0.45\columnwidth]{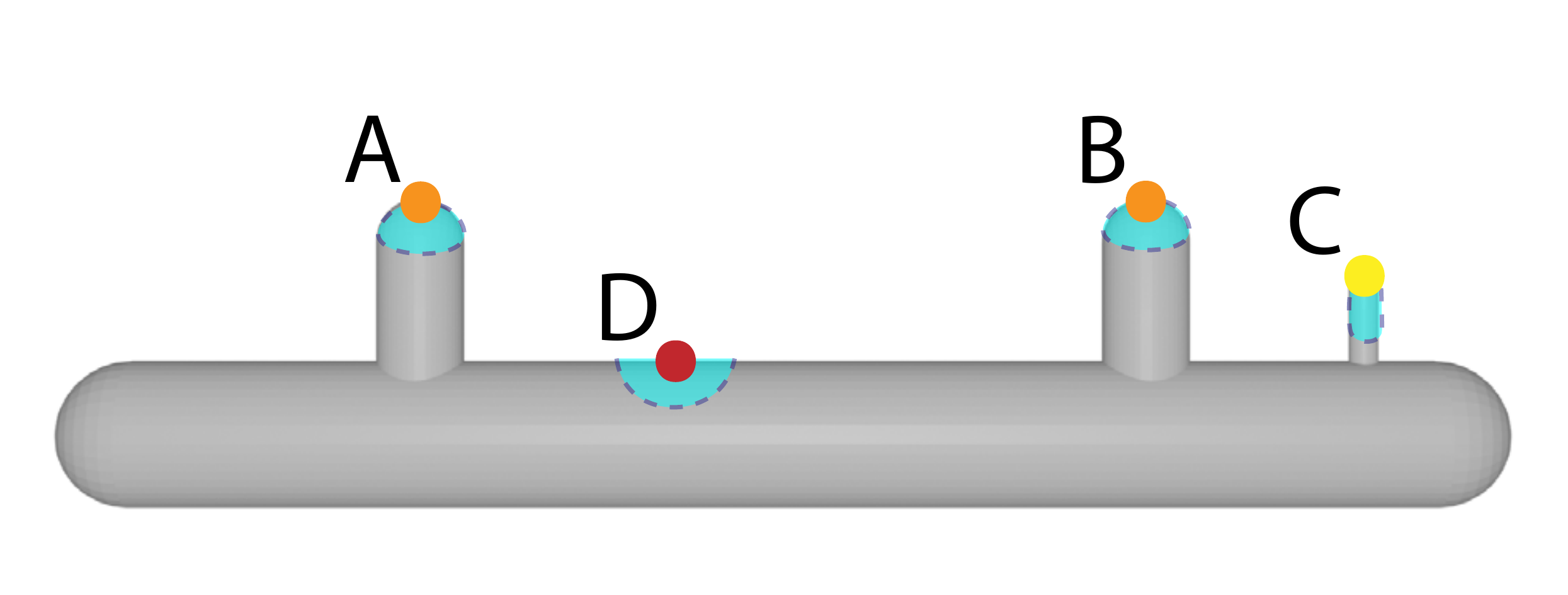}
  }
  \subfloat[\label{fig:heat_kernel_large}]{
    \includegraphics[width=0.45\columnwidth]{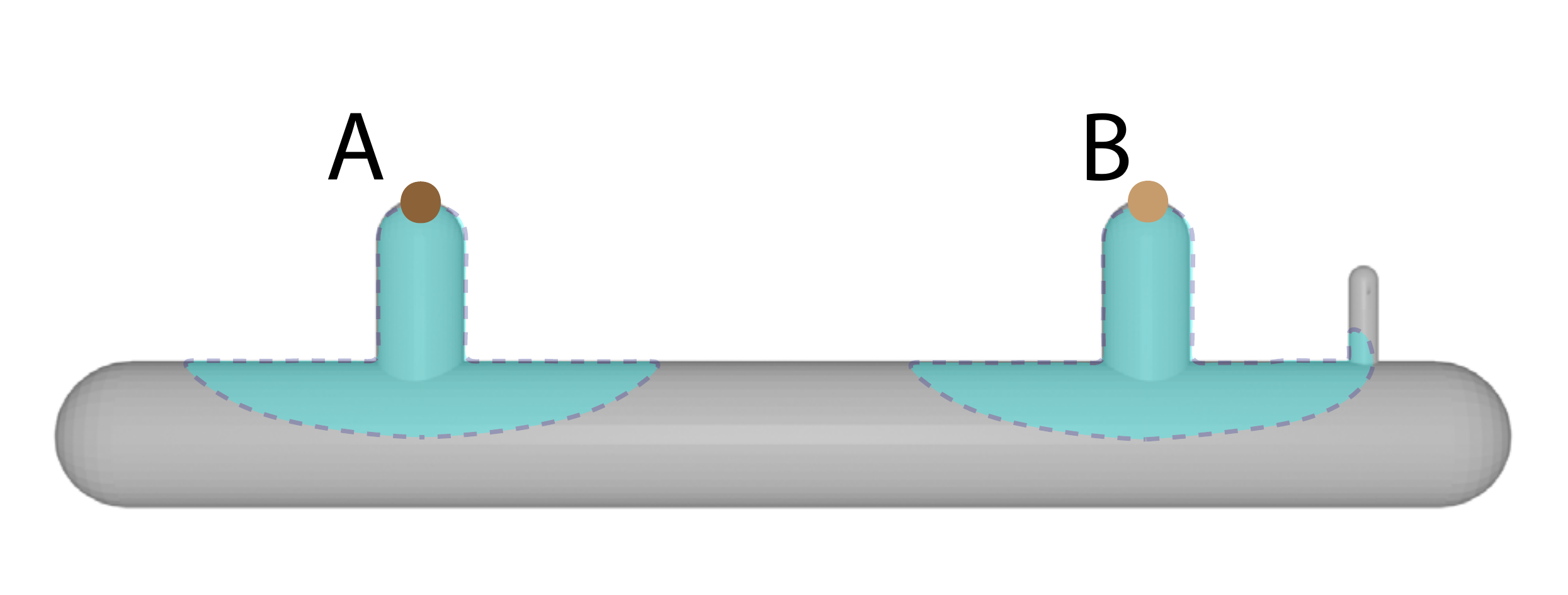}
  }
   \caption{An illustration of the HKS at two scales through examining the heat kernel $k_s(\mb{p}, \mb{p})$.  
The color of the points indicates the heat kernel at the specified scale, where the blue shaded region indicates the area that heat is allowed to diffuse.
For a smaller scale (a), the heat kernel captures mostly local geometric features (curvature on a 2D surface), while for a larger scale (b), the heat kernel  differentiates points by features in larger neighborhoods.\label{fig:heat_kernel}}
\end{figure}

It is often useful to restrict the heat kernel to look at ``a point to itself'' $k_s(\mb{p},\mb{p})$. This gives rise to the \emph{heat kernel signature} (HKS)~\cite{sun2009concise},
\begin{equation}\label{eq:hks_def}
	\textrm{HKS}(\mb{p}) = \{ k_{s_1}(\mb{p},\mb{p}) , k_{s_2}(\mb{p},\mb{p}) , \ldots , k_{s_{|S|}}(\mb{p},\mb{p}) \},
\end{equation}
sampled over a provided sequence of scales $S$ s.t.~$s_i \in S$.

To gain some intuition on what the HKS encodes, we consider a notional depiction on a 2D surface shown in \autoref{fig:heat_kernel}.
The HKS measures the amount of heat that is \emph{retained} at a given point across many scales. 
For small scales, heat diffuses faster in low curvature areas and slower in high curvature areas.
Thus we would get the highest HKS value for point $C$, lower for $A$ and $B$,
and $D$ will have the lowest HKS value.  
At larger scales, heat diffuses to larger areas and differentiates points by these larger neighborhoods.
For example, in \autoref{fig:heat_kernel_large}, the heat diffusion process from $B$ encounters the protrusion on the right, while diffusion from $A$ does not.  As a result, they would have different values for the HKS at this scale $s$.
In this way, the HKS gives us a multi-scale descriptor of the points that captures the intrinsic geometric properties.


\paragraph{Using the HKS for Visualization}

\begin{figure}[!t]
  \centering
    \includegraphics[width=1.0\columnwidth]{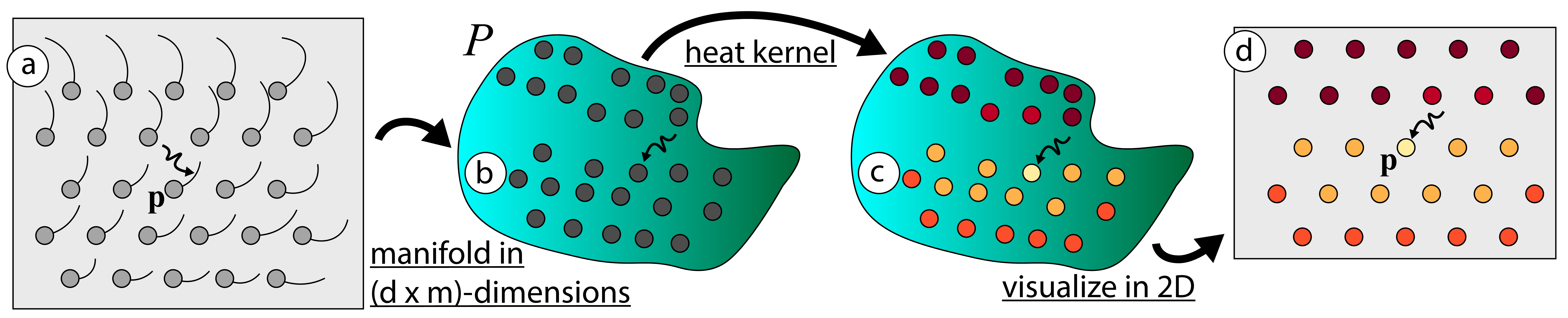}
	\caption{Overview of our approach. Given a collection of pathlines (a), we model each pathline as a point on a manifold living in the space of pathlines (b). We then construct the heat kernel on this manifold (c), and compute and visualize the HKS of the pathlines in the 2D domain (d).~\label{fig:overview}}
\end{figure}

We depict our procedure for computing the heat kernel and visualizing the HKS in Fig.~\ref{fig:overview}.  For unsteady flow represented as a set of pathlines (a), we first build our notion of shape by viewing the pathlines as points on a manifold (b), and then compute the heat kernel with respect to this manifold (c). Last, for visualization purposes, we compute the heat kernel signatures of the pathlines and visualize the HKS in the 2D domain (d). Each point in the 2D domain can be treated as a restriction of the full pathline geometry, and therefore each point in the 2D domain has a heat kernel signature associated with it. In this work, we visualize the HKS on the pathline's start position, $\mb{x}_{t_0}$.

Intuitively, when a pathline is similar to its neighboring pathlines on $P$, then heat may freely diffuse, resulting in a small amount of heat retained at that pathline. Conversely, if a pathline is different from its neighboring pathlines, then it will be more difficult for heat to diffuse, resulting in a large amount of heat retained. The HKS provides a multi-scale descriptor of this phenomenon, yet it is more than just a feature vector: one may interpret the HKS as a 1D function of scale, which can be useful for understanding how different types of flow behaviors manifest.

\begin{figure}[!t]
\centering
\subfloat[\label{fig:example1}]{
\includegraphics[width=0.49\columnwidth]{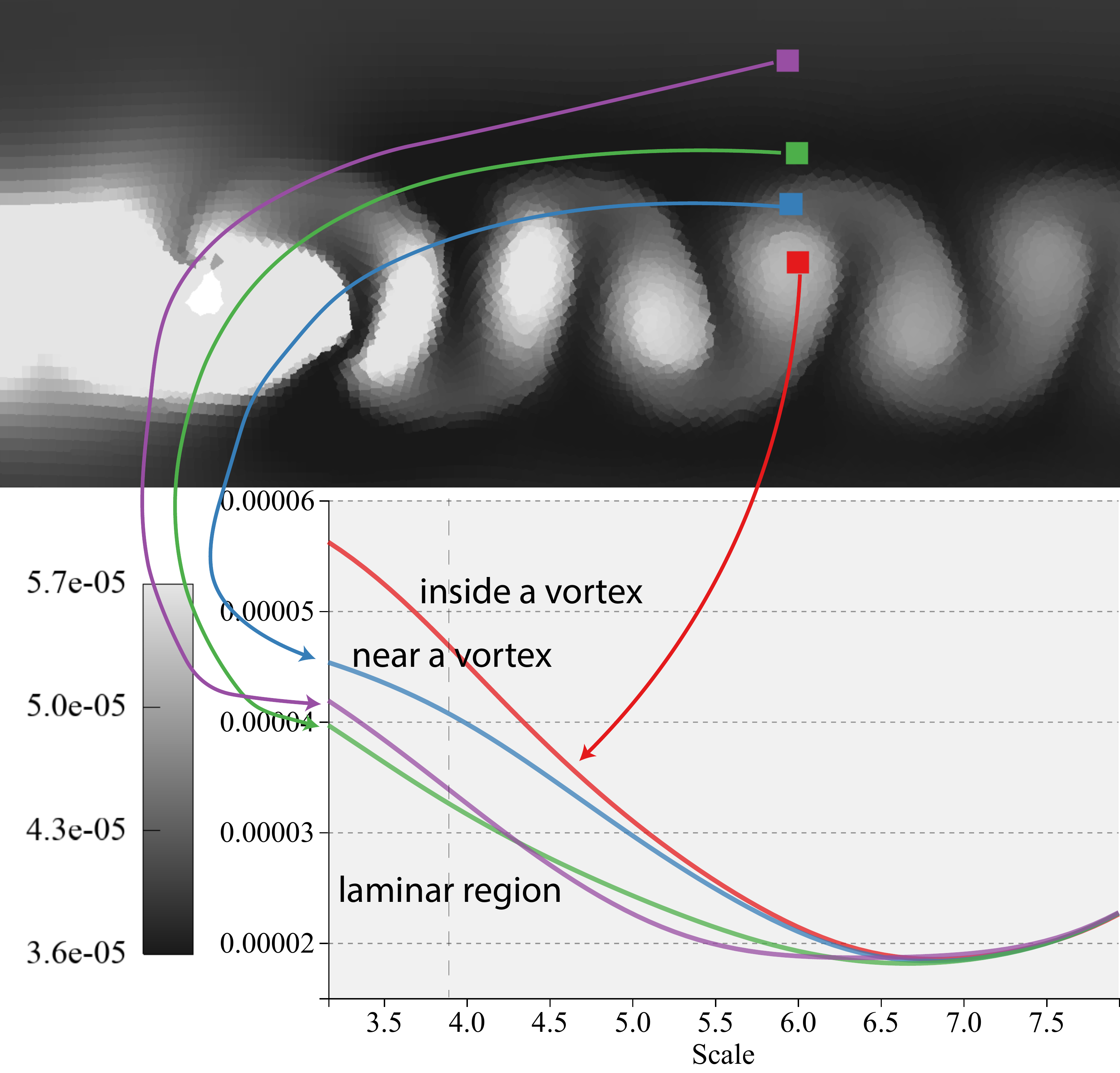}
}
\subfloat[\label{fig:example2}]{
\includegraphics[width=0.49\columnwidth]{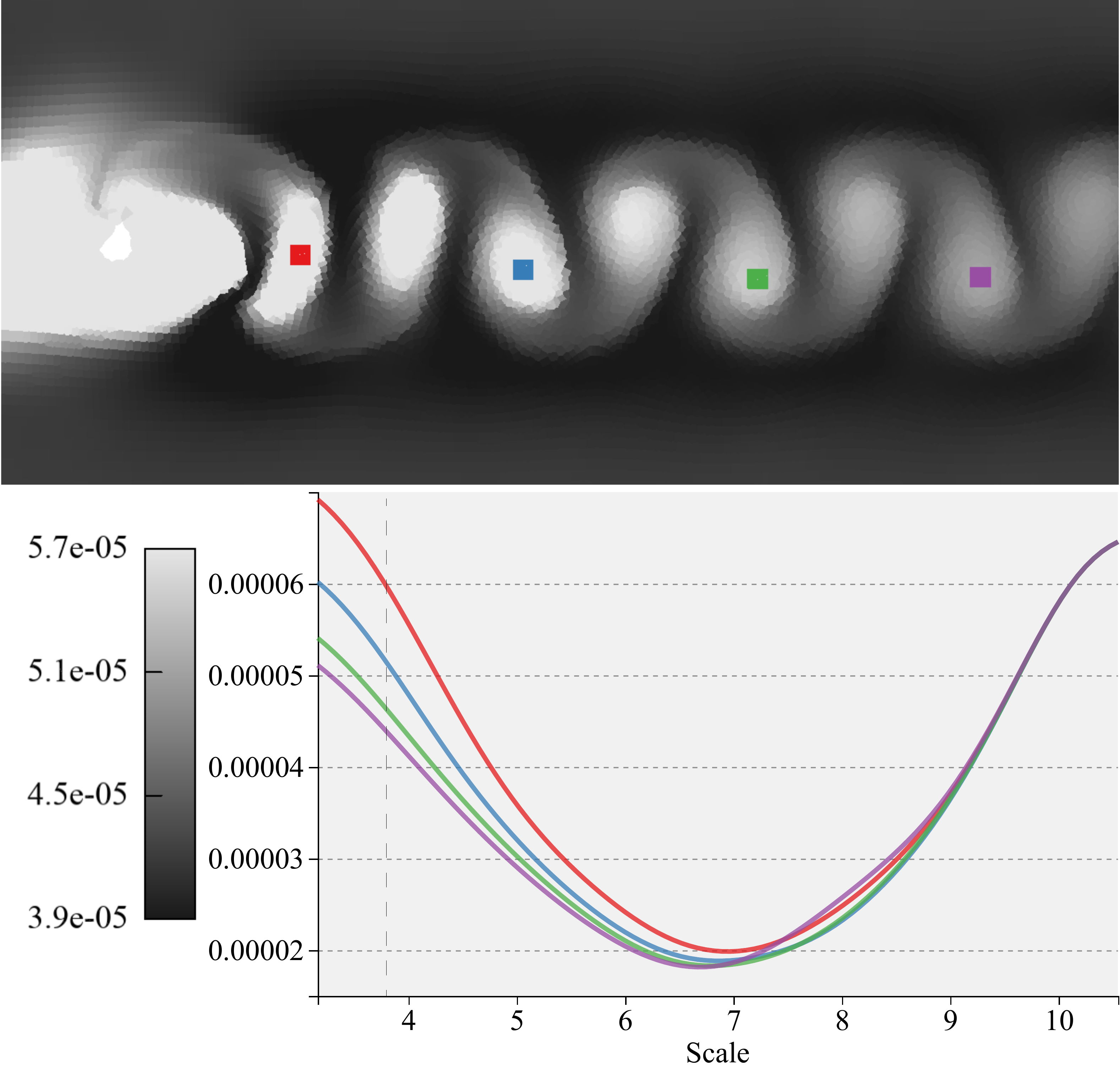}
}
\caption{In (a), we illustrate the HKS on flow over a cylinder with shedding vortices from left to right. On the top, we colormap the HKS at a specific scale, and show the full HKS of four pathlines on the bottom. The scale used on the top corresponds to the black vertical line on imposed on the bottom plot. The HKS enables us to discriminate distinct flow behaviors, as indicated by the labels in the plot.  In (b) we show the impact of vortex dissipation on the HKS, for flow over a cylinder. Here we see how the HKS can discriminate different strengths of vortices.}
\end{figure}

In Fig.~\ref{fig:example1} we illustrate the HKS for a dataset that models flow over a cylinder (described in more detail in Section~\ref{sec:results}). On top, we encode the HKS in the two-dimensional domain by showing the HKS at a specific scale as a greyscale value at the starting positions.  We also show HKS curves of the four colored pathlines on the bottom.  A pathline that lies in a vortex (red point) will only be similar to nearby pathlines of swirling motion, and it appears different from pathlines in laminar flow outside of the vortex.  Specifically, this pathline gives rise to an HKS that is large at all low-moderate scales suggesting that heat diffusion is limited to the vortex region. A pathline that lies near this vortex (blue point) will initially (at small scales) be able to freely diffuse to its neighborhood, but at a certain scale, the diffusion will hit the vortex region, thus limiting the heat diffusion, and giving rise to a larger HKS. This is analogous to the point labeled A in Fig.~\ref{fig:heat_kernel}, where in both cases, the spatial restriction of diffusion leads to a distinct HKS. Pathlines further in laminar regions (green and purple points) encounter no such barriers to diffusion at small or medium scales, but we can see the impact of the vortex on the green pathline's ability to diffuse, resulting in a slight larger HKS as scale increases. Since the HKS is commensurable, the impact of vortex vs. laminar vs. boundary regions can be observed even in different flow over cylinder datasets (shown in \autoref{fig:cylinder_two_timesteps}, and \autoref{fig:double}), enabling us to compare different flow and identify similar flow structures.

In Fig.~\ref{fig:example2} we show the HKS for the same flow over a cylinder dataset, but now highlighting pathlines that belong to different vortices. For this particular simulation, vortices are shed gradually dissipate, and we observe that the HKS curves are sensitive to this dissipation. Intuitively, vortex dissipation implies that the pathlines outside of the vortex gradually become more similar to pathlines within the vortex, enabling more heat to diffuse, and thus less heat retained, at the originating pathline. Note that at large scales, we see an inverse trend. At such large scales, the global geometry of the domain starts to impact diffusion, e.g. the pathline highlighted in purple will diffuse to the right boundary of the domain quicker than all of the other highlighted pathlines. Once diffusion hits the boundary, there is no place for heat to diffuse, and thus more heat is retained at the highlighted point. In contrast, the red pathline is closer to the center of the domain, and will be able to diffuse over more of the domain for larger scales.

\section{Methods}

We describe our method to compute and visualize the HKS values.

\subsection{Computing the HKS}\label{sec:compute_hks}

By the definition of the HKS (\autoref{eq:hks_def}) and the heat kernel's relation to the Laplace-Beltrami Operator (\autoref{eq:heat_kernel}), we can compute the HKS for a pathline $\mb{p}$ at a scale $s$ with
\begin{equation}
  \textrm{HKS}_s(\mb{p}) = k_s(\mb{p}, \mb{p}) = \sum_{i=0}^{m} e^{-\lambda_i s} \phi_i^2(\mb{p}),
\end{equation}
where $\lambda_i$ and $\phi_i$ are the eigenvalues and eigenvectors of the LBO $\Delta_P$ for manifold $P$.

As we have an unstructured sampling of $P$, with no connectivity, we adopt the approach of Liu et al.~(designed for point clouds) to compute a symmetrizable discrete approximation $\hat{L}_P^\sigma$ of $\Delta_P$~\cite{6264046}, and then perform eigendecomposation on $\hat{L}_P^\sigma$.  Note that we rename the parameter $t$ from that work as $\sigma$, to differentiate it from $t$ that we use for timestep of integration in the pathlines.  

The symmetrizable discrete approximation $\hat{L}_P^\sigma$ is basically an affinity matrix of the pathlines, approximating information about their spatial neighborhood on $P$.  We infer this spatial neighbor through their nearest neighbors computed with Euclidean distance.  Like Liu et al.~\cite{6264046}, we need to scale the affinity matrix by the ``local volume'' of each pathline, so that each pathline correctly represents a \emph{region} on the manifold rather than a point.
Local volume for a point $\mb{p}$ is computed by projecting $\mb{p}$'s neighbors to the tangent plane crossing $\mb{p}$, and then computing the volume of the Voronoi cell of $\mb{p}$ in the projected space.

\paragraph{Estimating Dimension of Local Neighborhood}

In the setting of Liu et al.~\cite{6264046}, they assumed they were working with two-dimensional manifolds.
Although we do not know the global dimension of $P$, we can, and only need to, estimate the \emph{local} dimension of each pathline so as to identify a reasonable target dimension to estimate local volume.  

We study this by using principal component analysis (PCA)  on a set of nearby pathlines and analyzing the eigenvalues of each principal component. For example, if the eigenvalue drops significantly from the second largest to the third largest, we can say that $\mb{p}$ is \emph{locally} 2-dimensional, meaning that its local neighborhood lives (mostly) on a 2D plane, even though each pathline is $d \times m$ dimensional, where $d$ is the dimension of the domain. Then we project the neighborhood of $\mb{p}$ to the 2-dimensional tangent plane and compute the volume of the Voronoi cell.  In practice, for each pathline $\mb{p}$, we consider its nearest neighborhood $N$ of size $M$, and compute PCA on $N$. We look at the eigenvalues associated to the principal components, and if the eigenvalues drop significantly from the $i$-th largest to the $i+1$-th largest, we say that $\mb{p}$ is locally $i$-dimensional.

We found that for our experiments with 2D flow datasets, nearly all pathlines (over 99\%) are locally 2-dimensional and thus this was a reasonable proxy for the volume of the local neighborhood. Interestingly, this experiment achieved the same result regardless of the number ($m$) of the pathline samples. That being said, the local dimensionality of pathlines is purely empirical in our current work.

\paragraph{Projecting the neighborhood and estimating volume}

For each pathline $\mb{p}$, we next gather its nearest neighborhood $N$ of size $M$, and project $N$ onto the tangent plane crossing $\mb{p}$ using PCA.  We can then compute the Voronoi cell $\Vor(\mb{p})$ of $\mb{p}$ on the projected plane and compute the volume of $\Vor(\mb{p})$ denoted as $\textrm{vol}(\Vor(\mb{p}))$. In practice, we use Qhull\cite{barber1996quickhull} to compute the Voronoi diagram as well as $\textrm{vol}(\Vor(\mb{p}))$. Since $P$ is likely a manifold with boundary, we need to be especially careful with the pathlines near the boundaries of the manifold. Specifically, after we have computed the Voronoi cell of $\mb{p}$, we check if there is any Voronoi vertex in the cell that is outside of the projected neighborhood of $\mb{p}$. If so, this suggests that $\mb{p}$ is near boundary, we add 4 extra points around $\mb{p}$ aligned with the direction of the principal components at the same distance as the nearest neighbor of $\mb{p}$. This way we bound the Voronoi cell of $\mb{p}$ by its distance to its nearest neighbor, if only $\mb{p}$ is near the boundary of $P$ and its Voronoi cell potentially is unbounded.

In our experiments with 2D flow data, we typically have less than 10\% of pathlines that are near the boundaries of $P$. Most of these pathlines live near the spatial boundaries of the flow domain, while some of them live near highly turbulent areas (e.g. the region immediately behind the cylinder in the flow over cylinder dataset.)

\paragraph{Assembling the discrete LBO and Eigendecomposition}

We assemble the discrete LBO matrix $\hat{L}^\sigma_P = B^{-1}\cdot Q$, where $B$ is a diagonal matrix and $Q$ is symmetric. We can compute elements $q_{ij}$ and diagonal elements $b_{ii}$ as the following:
\begin{equation}
  q_{ij} = \textrm{vol}(\Vor(\mb{p}_i))\textrm{vol}(\Vor(\mb{p}_j))\frac{1}{4\pi{}\sigma^2}e^{-\frac{\norm{\mb{p}_i - \mb{p}_j}^2}{4\sigma}},
\end{equation}
where $i \neq j$, $\mb{p}_j \in N_{\mb{p}_i}$, $\sigma$ attenuates how we weight the neighborhood of each point, and 
\begin{equation}
  q_{ii} = -\sum_{j \neq i}{q_{ij}}, \textrm{and}
\end{equation}
\begin{equation}
  b_{ii} = \textrm{vol}(\Vor(\mb{p}_i)).
\end{equation}



After we assembled discrete approximation of LBO $\hat{L}^\sigma_P$, we symmetrize it with 
\begin{equation}
  U = B^{\frac{1}{2}} \cdot \hat{L}^\sigma_P \cdot B^{-\frac{1}{2}} = B^{\frac{1}{2}} \cdot B^{-1} \cdot Q \cdot B^{-\frac{1}{2}}
\end{equation}

Then we use a sparse eigendecomposition routine to compute eigenvalues $\hat{\lambda_i}$ and eigenvectors $\hat{\phi_i}$ for $U$, and we get

\begin{equation}
  \lambda_i = \hat{\lambda_i}, \phi_i = B^{-\frac{1}{2}} \cdot \hat{\phi_i}
\end{equation}

Lastly, we use eigenvalues $\lambda_i$ and eigenvectors $\phi_i$ to compute the HKS for every point. Note that our effective range of scale is bounded by the eigenvalues. That is, we have
\begin{equation}
  s_{min} = -\frac{\log{\beta}}{\lambda_{min}}, s_{max} = -\frac{\log{\beta}}{\lambda_{max}},
\end{equation}
where $\beta$ is a threshold of precision set by the user. In practice we set it to be $0.01$. Also, the smallest eigenvalue will always be $0$, therefore we use the second smallest eigenvalue to compute $s_{min}$. We use log scale to sample $100$ scales from $[s_{min}, s_{max}]$, and we normalize the HKS following the practice of Sun et al.~\cite{sun2009concise}.


\subsection{Visual Interface}

Our visual interface enables the user to navigate and compare timesteps of one or two datasets with various perspectives powered by the HKS. \autoref{fig:teaser} shows an overview of our visual interface and its four major views, namely the Mean HKS view, Single-Scale HKS view, Point Similarity view, and HKS Cluster view. 

\paragraph{Components of the Interface}

On the left of \autoref{fig:teaser} is an overview of our visual interface. The top left is the HKS viewer where the selected HKS view is rendered. There are two viewports in the HKS viewer corresponding to two timesteps selected by the user using the sliders below. On the right are the rendering options, where the user can toggle trajectory and point rendering, select which HKS view to show on the HKS viewer, and commence computation of clusters. On the bottom of the interface are options for colormapping and selecting HKS scale range for visualization.


Selecting and visualizing different HKS views are the crucial part of analyzing flow datasets using our visual interface. Each HKS view has its purpose of presenting the HKS and enabling the comparison and understanding of the flow.  To show the HKS, we compute a Voronoi diagram in the ambient 2D space of the flow dataset, and then color map values to determine what color to draw the Voronoi cell with.  This creates a visual interpolation of the values in question.  For speed, we use a shader-based approach that draws cones, and we let the depth buffer resolve the boundaries of the Voronoi edges.

\paragraph{Mean HKS View}

The Mean HKS view visualizes the average of HKS of selected scale range for each pathline in the timesteps. This view is intended to present an overview of the flow at this timestep. When this view is enabled, the scale range selection sliders in the HKS View Options select the maximum scale for the average HKS for each timestep. The minimum scale for computing the average HKS is set to be the minimum scale of the HKS of pathlines in this timestep. In \autoref{fig:teaser} for example, the scale ranges for the left and right timesteps are set to be the same, enabling the HKS viewer to present a comparison of overviews of the two timesteps.

\paragraph{Single-Scale HKS View}

The Single-Scale HKS view visualizes the HKS of pathlines in the timesteps at a single scale selected using the scale selection sliders. This view is mainly designed to help understand what feature of the flow is captured by the HKS at individual scales. In the example of Single-Scale HKS view in \autoref{fig:teaser}, a smaller scale is selected for the left viewport while a larger scale is selected for the right. The result verifies that smaller scales of the HKS capture more localized features while larger scales captures more global features.

\paragraph{Point Similarity View}

In the Point Similarity view the user can select a point (pathline) in the HKS viewer, and the viewer will visualize the distances of all other points in both timesteps to the selected point with regard to the HKS of selected range. This view enables the analysis of individual flow features or patterns of interest within the context of either its own timestep or the comparison of another.  For example, in the Point Similarity view shown in \autoref{fig:teaser}, a point in the top right vortex of the left timestep is selected and other vortices and turbulent regions in both timesteps are highlighted in the viewer. Vortices that are similar with the selected one in size is closer to it in terms of the HKS, like the one next to it. Vortices that are different in size would be further in terms of the HKS, like the vortices in the other timestep which is later in time, causing the vortices to dissipate more.

When this view is enabled, the HKS scale sliders select the range of scale that's been used to compute the distances. From our experiments, fixing the minimum of the scale range to be the minimum scale of the HKS of the pathlines and varying the maximum of the scale range yields reasonable visualizations of comparisons for various flow features and patterns.

\paragraph{HKS Curves}
When selecting a point in either of the three views mentioned above, the HKS curve of the point will be presented in a separate window for direct comparisons of HKS curves. For the purpose of easy referencing, the curves and points will have matching colors, and points in different timesteps will have different curve stroke types (solid vs.~dotted).

\paragraph{HKS Clusters View}

In HKS Clustering view, the Mean HKS view will first show in the viewer as an overview of the HKS. Next, the user can select a rectangle region for both timesteps, and push the compute button on the right to compute the clusters for points in the selected regions with respect to their HKS either jointly or separately for each timestep. When the clusters are computed, they will be visualized in the viewer, and the HKS curves of the centroids of the clusters will be presented in a separate window.  The user can selected the scale range of the HKS for the clusters as well as the number of clusters in the HKS View Options.  Using the HKS Clustering view, different regions of flow will be distinguished, while recurrent or symmetric patterns can be identified by the clusters.

\subsection{Discussion of Parameters}

As described in \autoref{sec:compute_hks}, the user needs to set a few parameters when computing the HKS. These parameters are important for the HKS to be useful for comparison and analysis. 

The number of neighbors for each point $M$ needs to be set large enough to have a good approximation of the local neighborhood on the manifold, but it should not be set too large, or it will span holes in the manifold. In our experiments we typically set $M=30$ to get reasonable results.

When assembling the symmetrizable discrete LBO $\hat{L}_P^\sigma$, we do not check if $\mb{p}_j$ is inside $\mb{p}_i$'s neighborhood in practice. 
Instead, we compute $q_{ij}$ for all $i$ and $j$ and zero out all $q_{ij} < T$, where $T$ is the threshold set manually to control the sparsity of $\hat{L}^\sigma_P$. From our experiments, varying $T$ did not have a noticeable effect on the HKS for the points. However, it does affect the computation time for eigendecomposition. Typically, setting $T$ so that the percentage of non-zero entries per row for $\hat{L}^\sigma_P$ falls in $[1\%, 10\%]$ will result in reasonable performance of eigendecomposition. To make this approach scalable for larger datasets, one can use a nearest neighbor data structure to filter out points that are too far.

As for the parameter $\sigma$, it should be set relative to the sample density of the point cloud with respect to the manifold. We compute the maximum distance between a point and its neighbor $\delta_\mb{p} = \max_{\mb{q} \in N}{\norm{\mb{p} - \mb{q}}}$. Then we compute $\eta$ as the median of $\delta_\mb{p}$ for all point $\mb{p} \in P$.  We model $\sigma$ as 
\begin{equation}\label{eq:diffusion_scale}
  \sigma = \alpha * \eta,
\end{equation}
where $\alpha$ is the parameter we need to set separately for each dataset. From our experiments, $\alpha = 0.01$ is typically a good starting point. If we increase $\alpha$, we essentially connect each point to a larger neighborhood, allowing the heat to diffuse more. We may lose some local features by doing this. If we decrease $\alpha$, we restrict the heat diffusion on the manifold, allowing more local features to be captured. However, if $\alpha$ is set to be too small, the discrete LBO would be too sparse and eigendecomposation would fail.

\begin{figure}[!ht]
  \centering
  \subfloat[]{
    \includegraphics[width=0.21\columnwidth]{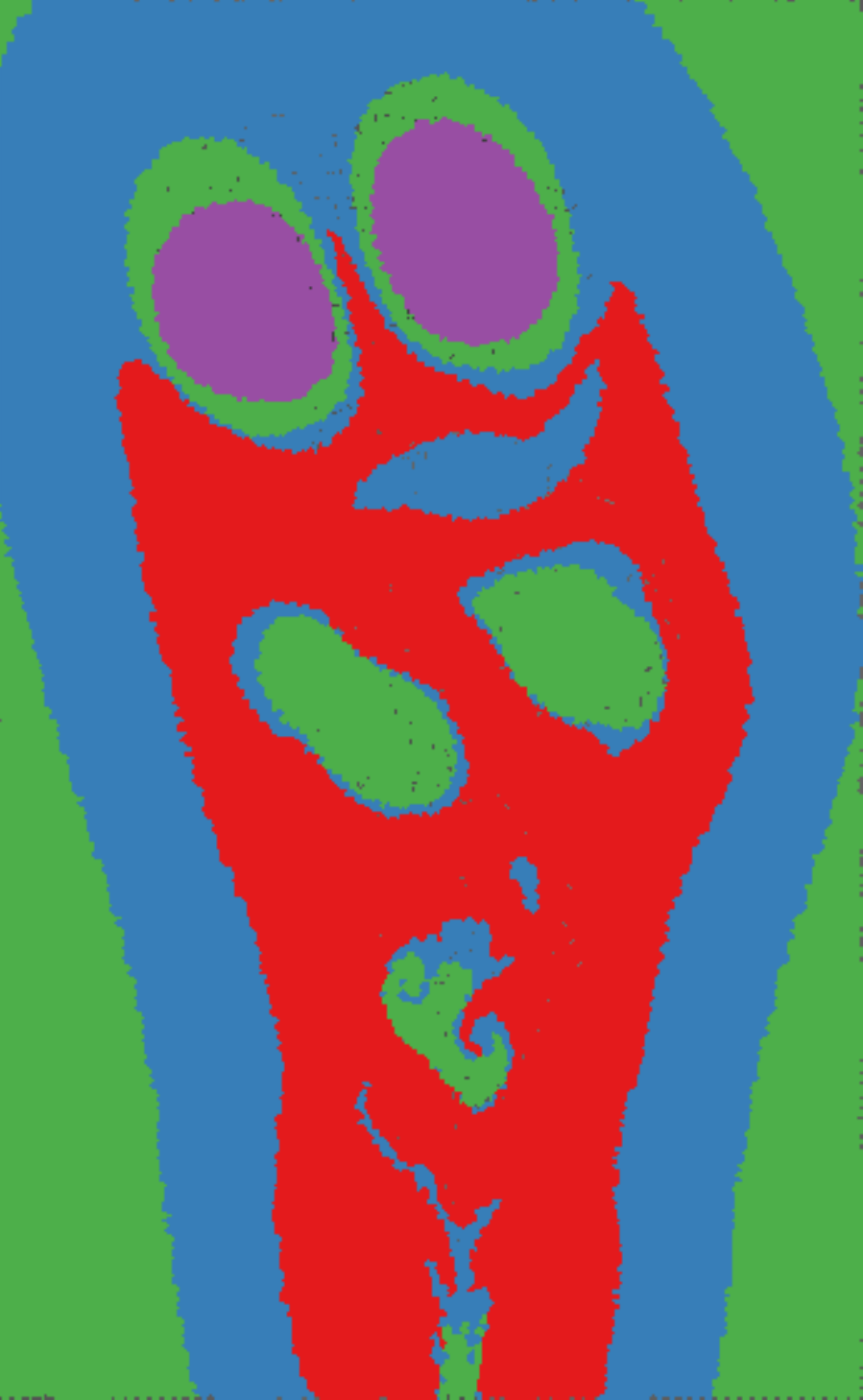}
  }
  \subfloat[]{
    \includegraphics[width=0.21\columnwidth]{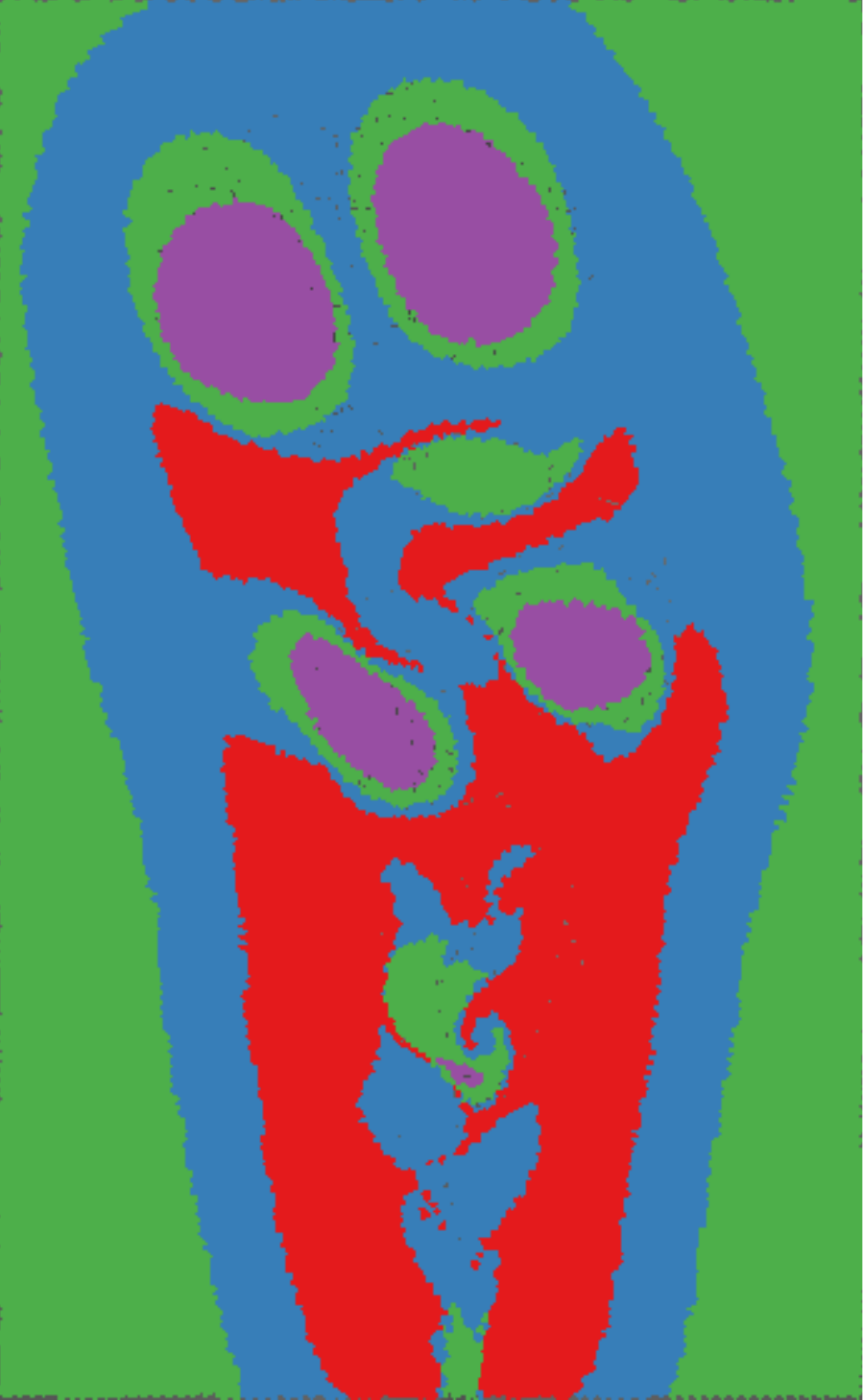}
  }
  \subfloat[]{
    \includegraphics[width=0.21\columnwidth]{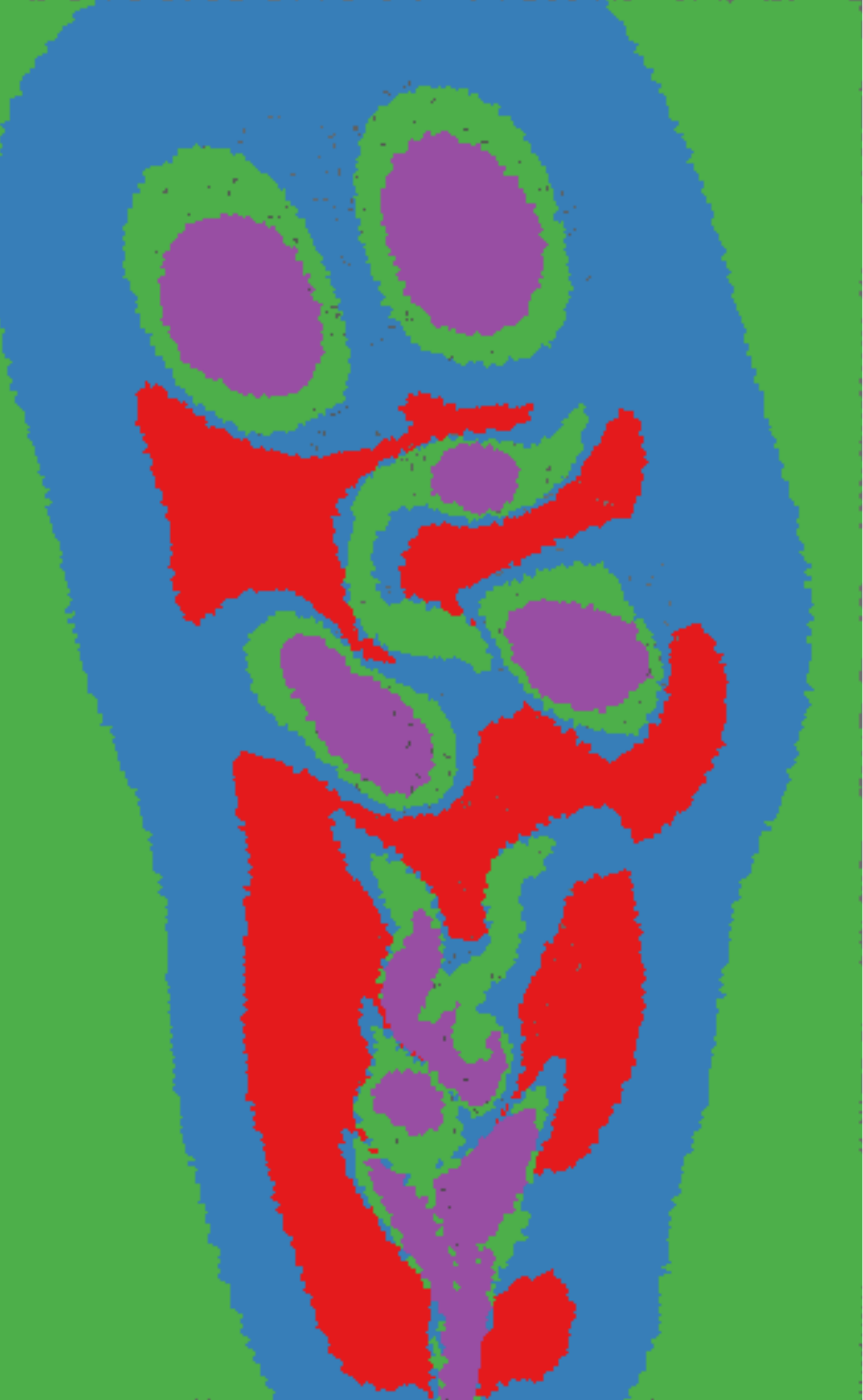}
  }
  \subfloat[]{
    \includegraphics[width=0.21\columnwidth]{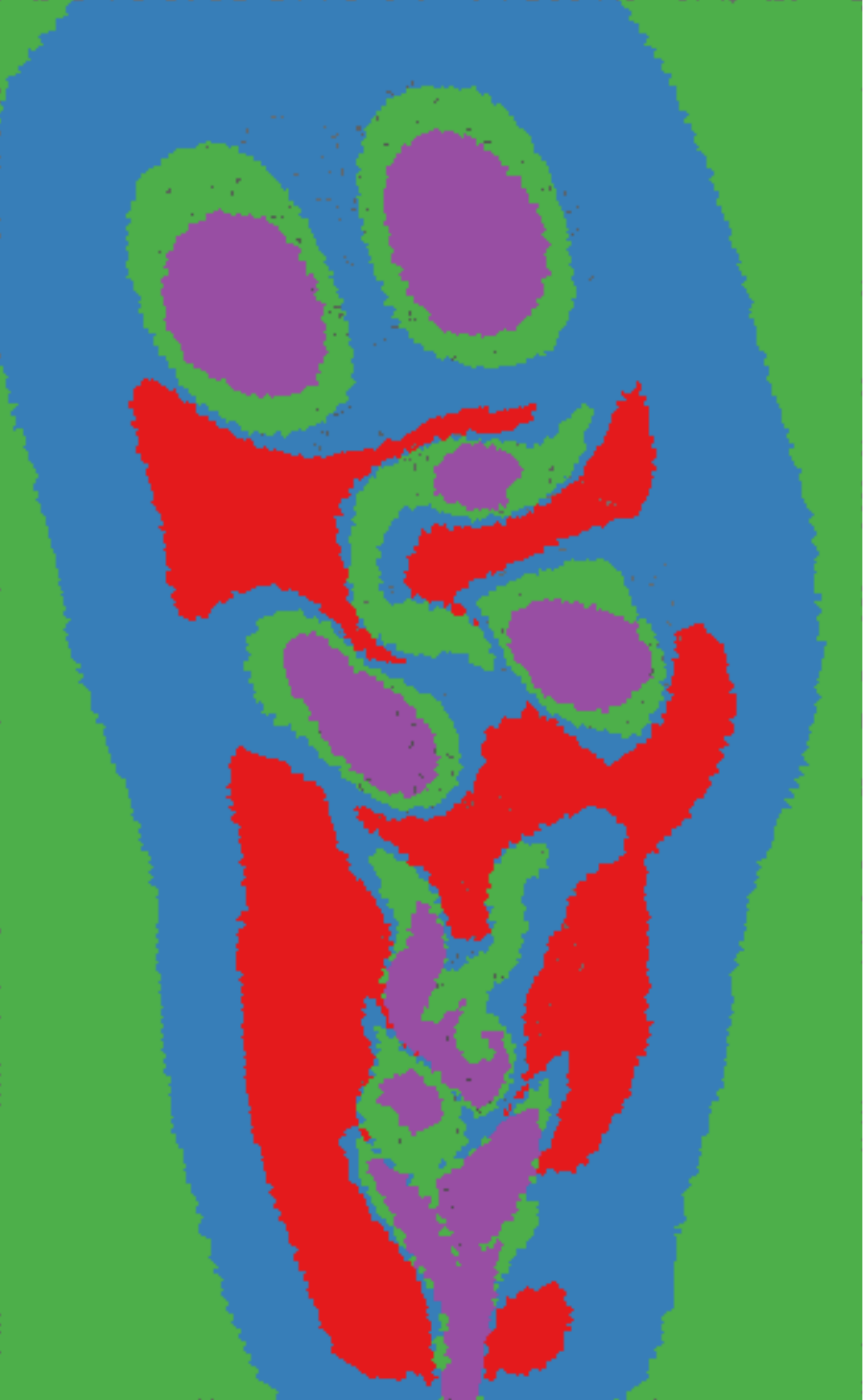}
  }
  
  \caption{\label{fig:heated_numofsample} Clusters of a heated cylinder convected flow
  dataset with different number of samples $m$ for each pathline. All timesteps are uniformly sampled.
  number of samples $m$: (a) 2 (b) 3 (c) 17 (d) 50}
\end{figure}

Lastly, although the number of samples $m$ for each pathline is not a parameter to be set for computing the HKS and the visual interface, it is worth discussing its influence on the HKS of the pathlines. Intuitively, the more samples we have, the more ``accurate'' the pathlines are and the more discriminating power the HKS will have. \autoref{fig:heated_numofsample} shows clustering results of a heated cylinder convected flow dataset with varying number of samples for the pathlines. We can see that from $2$ samples to $3$ and to $17$ samples, the clustering results change dramatically, whereas from $17$ to $50$ samples, the results barely change. This indicates that our analysis with the HKS converges given enough samples for the pathlines.

\section{Results}
\label{sec:results}


We have applied our technique to datasets that are analytic equations as well as simulation outputs. We precompute the HKS of the datasets before analyzing them in our visual interface. The statistics, running time, and parameters of the datasets can be seen in Table \ref{tab:results}.

\begin{table*}[!ht]
  \caption{\label{tab:results}Statistics, running time, and parameters used for the datasets. Computing the HKS for different starting timesteps of the same flow or each of the heated cylinder flow ensemble has similar running time, therefore information for one typical configuration is listed for each flow. For the Unsteady ABC flow, we pick one full cycle to compute the HKS. The last four columns are times, in seconds, of estimating volumes of Voronoi cells for all points, assembling the Laplace-Beltrami Operator, computing the eigendecomposition, computing the HKS.}
  \centering
  \begin{tabular}{|l|l|l|l|l|l|l|l|l|l|}
    \hline
    name                & num\_pathlines & $t_0$  & $\tau$ & $m$ & $\alpha$ & Volumes & LBO    & eigendecomposition & the HKS  \\ \hline
    unsteady ABC        & 40000          & 0    & /      & 30 & 0.5      & 57.62   & 130.29  & 134.30             & 0.27 \\ \hline
    single 2D cylinder  & 57525          & 55   & 5.4    & 150 &0.1     & 1071.24   & 1587.02  & 809.13             & 0.45 \\ \hline
    double 2D cylinders & 61753          & 55   & 5.4    & 150 &0.1     & 1152.17   & 1826.49  & 872.95             & 0.45 \\ \hline
    heated cylinder $A_1$    & 161093         & 6.86 & 0.85 & 50 & 0.03     & 279.73   & 4863.69 & 2339.88            & 1.09 \\ \hline
    \end{tabular}

\end{table*}


\paragraph{Unsteady ABC Flow}

\begin{figure}[!ht]
  \centering
  \subfloat[\label{fig:abc_a}]{
    \includegraphics[width=0.23\columnwidth]{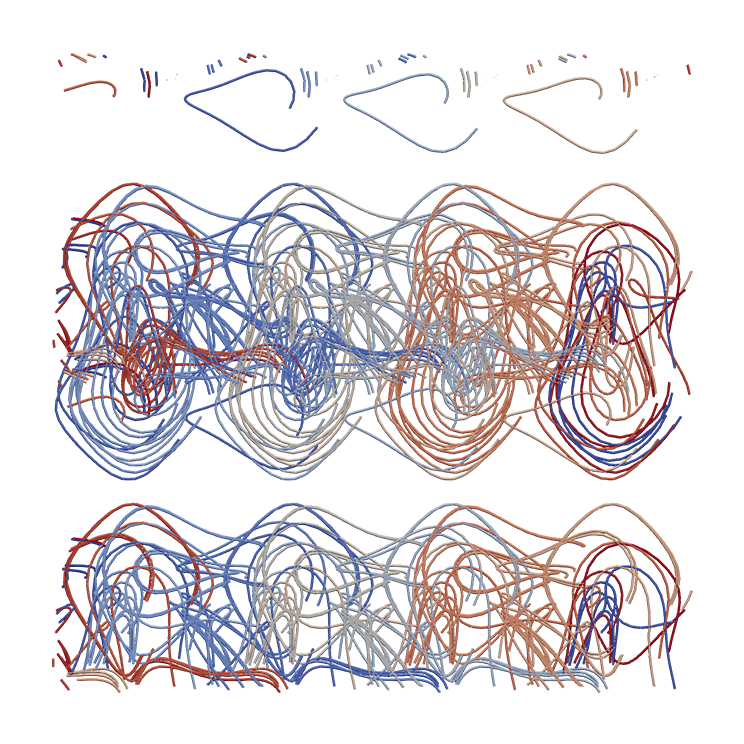}
  }
  \subfloat[\label{fig:abc_b}]{
    \includegraphics[width=0.23\columnwidth]{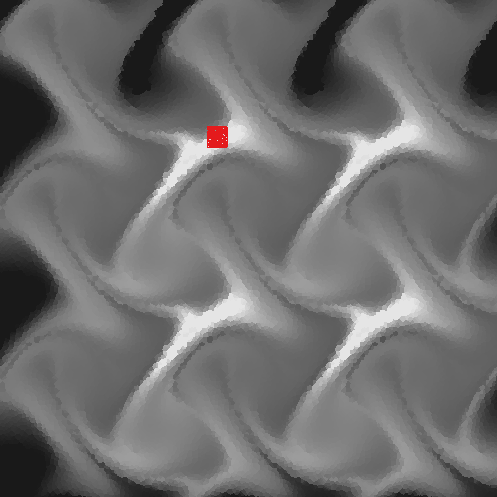}
  }
  \subfloat[\label{fig:abc_c}]{
    \includegraphics[width=0.23\columnwidth]{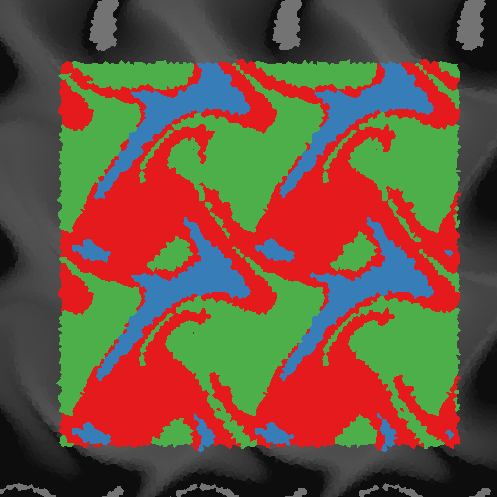}
  }
  \subfloat[\label{fig:abc_d}]{
    \includegraphics[width=0.23\columnwidth]{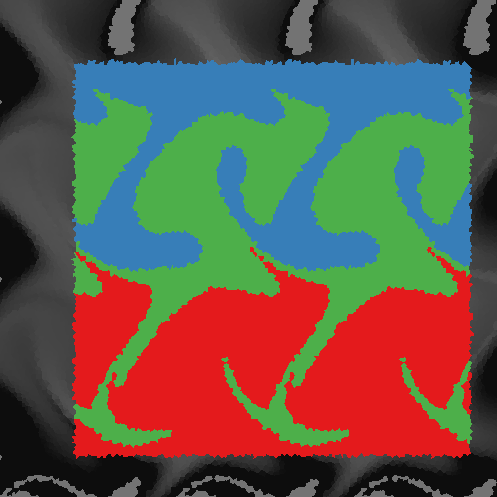}
  }
  \caption{The HKS can capture intrinsic symmetric features of unsteady flow. (a) A subset of the input pathlines, colored by pathline id. (b) Point Similarity  with respect to the red selected point. (c) HKS cluster view with the same scale range and 3 clusters. (d) Pathlines clustered by their raw positions.  Seed position dominates the clustering resulting in groups that  partially express recurrent flow but fail to group symmetries.\label{fig:abc}}
\end{figure}

We first analyze a simple recurrent flow generated by the analytic equations for the unsteady variant of the periodic 3D flow known as the ABC (Arnold-Beltrami-Childress) Flow.  Our modification is based on replacing $z$ with the time dimension, as done by Shi et al.~\cite{shi2006path}.  This leads to the following equations for the vector field:
\begin{equation}
	\mb{v}(x,y,t) = \begin{pmatrix} A \sin(t) + C \cos(y) \\
  B \sin(x) + A \cos(t) \end{pmatrix}
\end{equation}
with the standard parameters $A=\sqrt{3}$, $B=\sqrt{2}$, and $C=1$.
We consider pathlines that are uniformly-seeded in the domain $[0,8\pi]^2$ and integrate pathlines starting at $t_0 = 0$ for duration $\tau$ of a full temporal cycle ($\tau = 2\pi$). For each pathline, we use $m=30$ uniformly distributed sample timesteps. A coarse subset of the pathlines is shown in \autoref{fig:abc_a}. By characterizing the intrinsic symmetries in the pathline manifold, the HKS can capture the symmetric patterns of the flow. With the HKS and our tool, we can highlight the symmetric features of this dataset using Point Similarity view as shown in \autoref{fig:abc_b}. Using k-means clustering with $k=3$, we can differentiate regions of a recurrent pattern (\autoref{fig:abc_d}). By comparison, \autoref{fig:abc_d} shows the pathlines clustered using their raw positions, which partially distinguishes recurrent flow, but fails to group repeated patterns. Our method thus produces qualitatively distinct behavior compared to other geometry-based clustering methods~\cite{hadjighasem2016spectral,banisch2017understanding}.


\begin{figure}[!ht]
  \centering
  \subfloat[\label{fig:cylinder_single_cluster_2}]{
    \includegraphics[width=0.49\columnwidth]{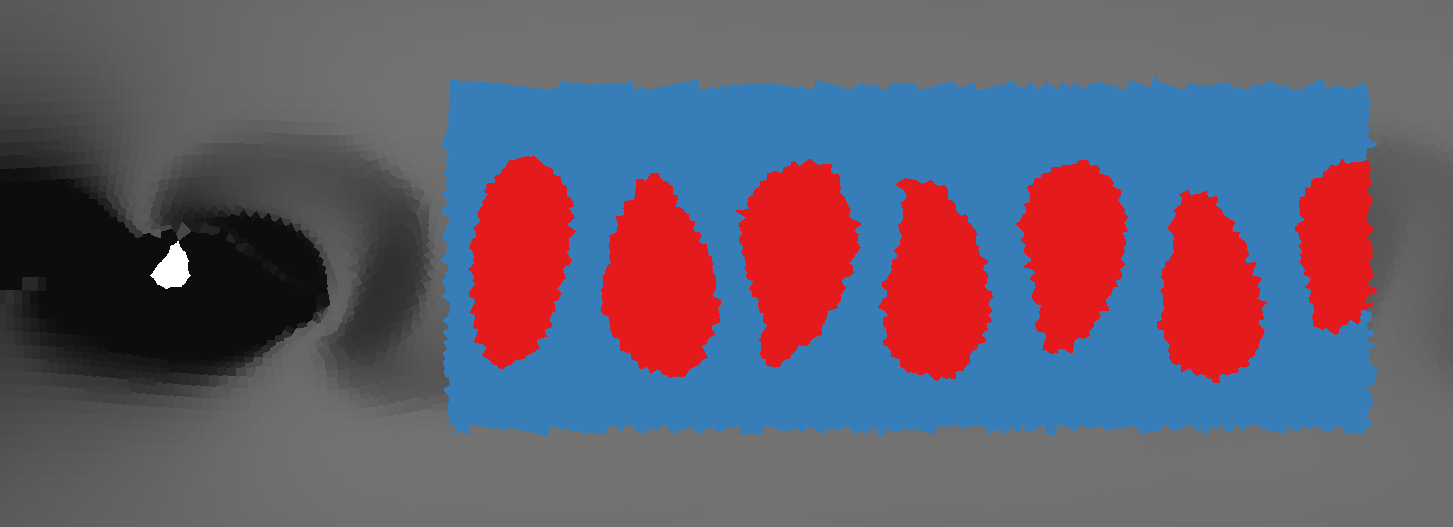}
  }
  \subfloat[\label{fig:cylinder_single_cluster_4}]{
    \includegraphics[width=0.49\columnwidth]{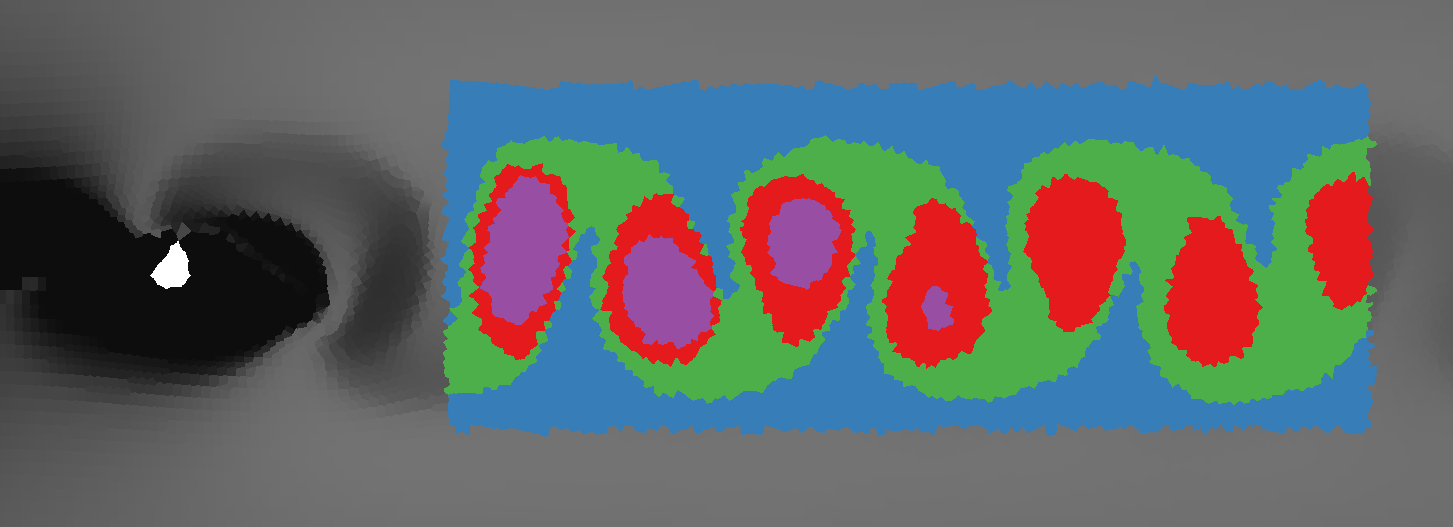}
  }
   \caption{We show the HKS Cluster views of the vortex street with different number of clusters.\label{fig:cylinder_single_cluster}}
\end{figure}

\paragraph{Flow over 2D Cylinder}

We further analyze the flow over cylinder dataset we have looked at in \autoref{fig:example1} and \autoref{fig:example2}. The simulation is generated by fluid simulation software Gerris~\cite{popinet2003gerris} on a domain of $[-0.5,15.5]\times[-3.5,3.5]$. The flow goes from left to right with initial velocity of $1$ and viscosity set to $0.00078125$. A cylinder of radius $0.125$ is centered at $(4, 0)$. The result of the simulation is a von K\'{a}rm\'{a}n vortex street that forms in the wake of the cylinder as the Reynolds number of the configuration is 160. We then extract pathlines from the time-varying vector fields and compute the HKS. We consider pathlines that start at two different times $t_0^0 = 55$ and $t_0^1 = 56.5$ with integration duration of $\tau=5.4$. For each pathline, we use $m=150$ uniformly distributed sample timesteps.

We have analyzed the flow over cylinder at a single timestep with \autoref{fig:example1} and \autoref{fig:example2}. We showed that using the HKS, we can differentiate vortex and laminar flow, as well as dissipating vortices. We confirm this result with \autoref{fig:cylinder_single_cluster}, which shows k-means clustering result on the region near the vortex street. \autoref{fig:cylinder_single_cluster_2} shows clustering with $k=2$ which separates the vortex cores from the background flow.  But, when clustering with $k=4$ (\autoref{fig:cylinder_single_cluster_4}) we see that pathlines from the early clusters are distinguished as we move from left-to-right.

\begin{figure}[!ht]
  \centering

  \subfloat[\label{fig:double_a}]{
    \includegraphics[width=0.7\columnwidth]{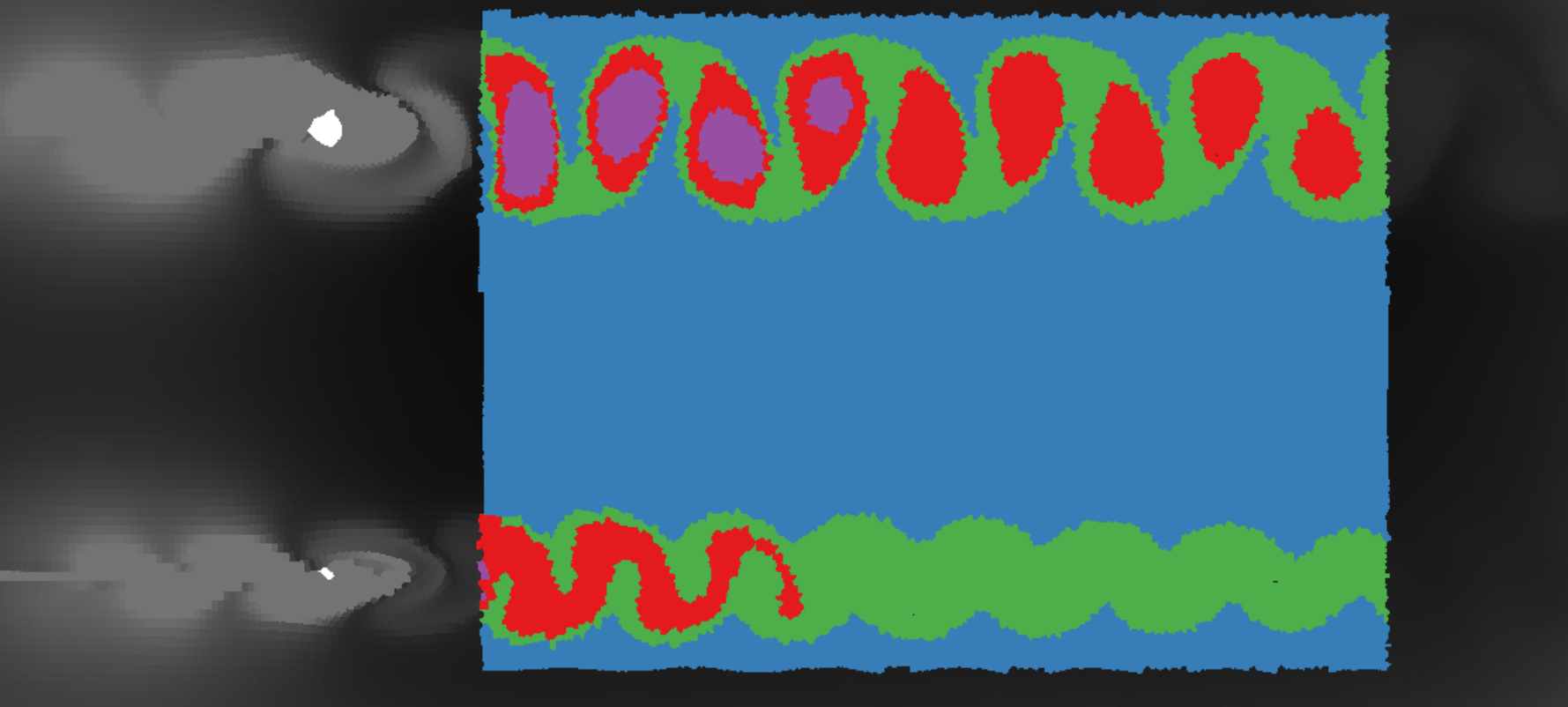}
  }

  \subfloat[\label{fig:double_b}]{
    \includegraphics[width=.9\columnwidth]{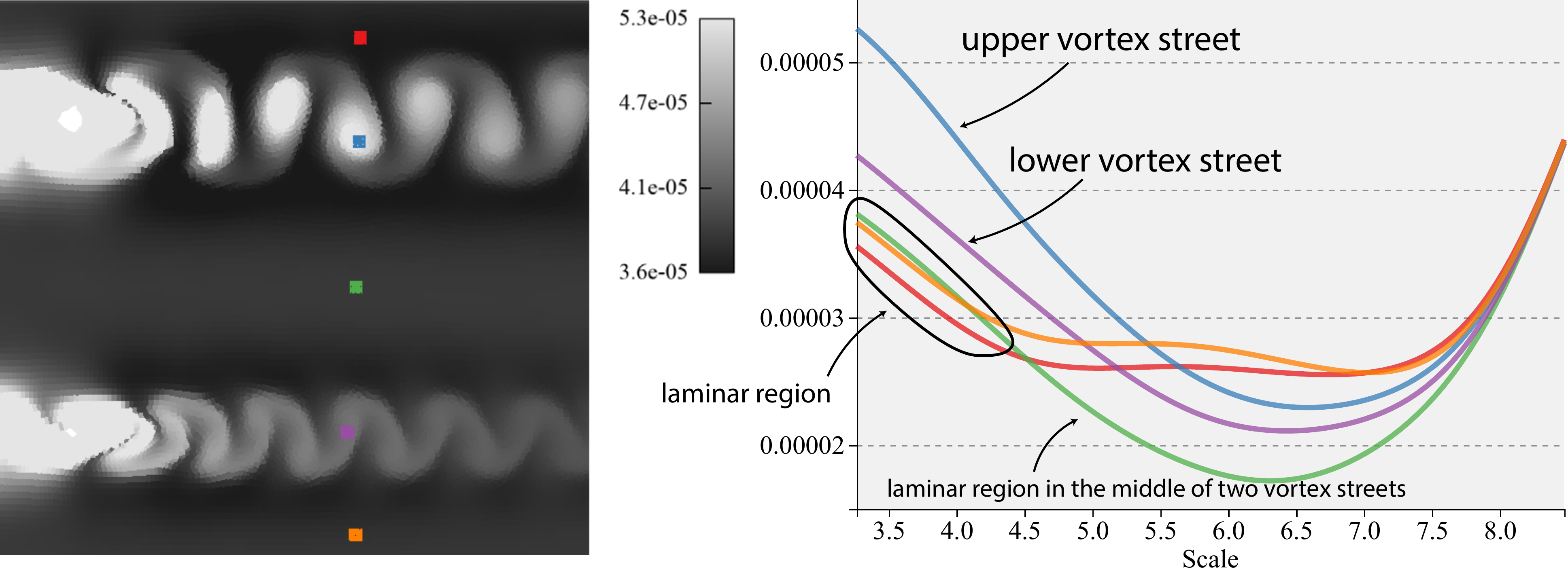}
  }
   \caption{\label{fig:double} (a) Flow over double cylinders using k-means clustering with $k=4$. (b) Five points with their HKS curves.}
 \end{figure}

\begin{figure}[!ht]
  \centering
  \subfloat[\label{fig:cylinder_two_timesteps_a}]{
    \includegraphics[width=0.45\columnwidth]{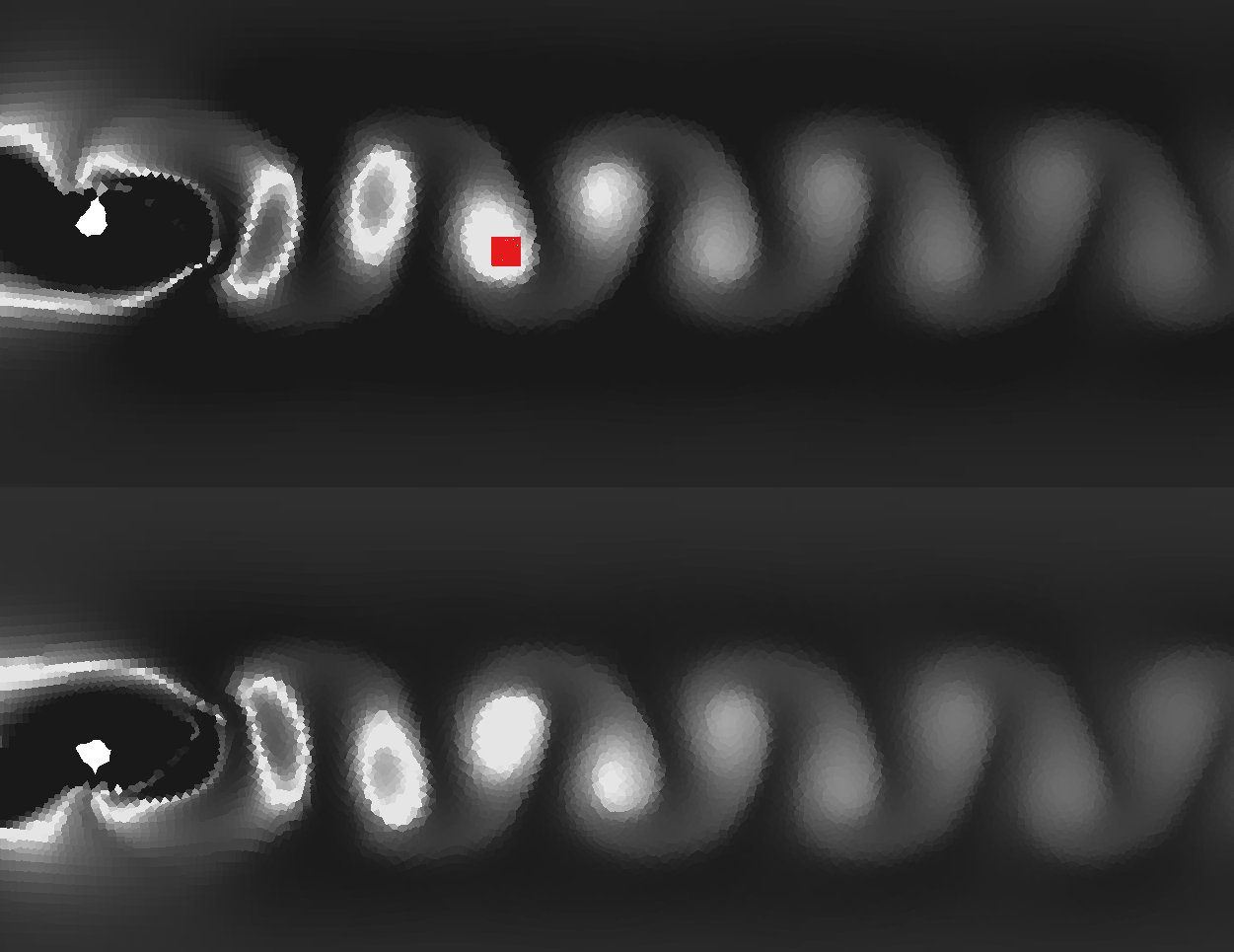}
  }
  \subfloat[\label{fig:cylinder_two_timesteps_b}]{
    \includegraphics[width=0.45\columnwidth]{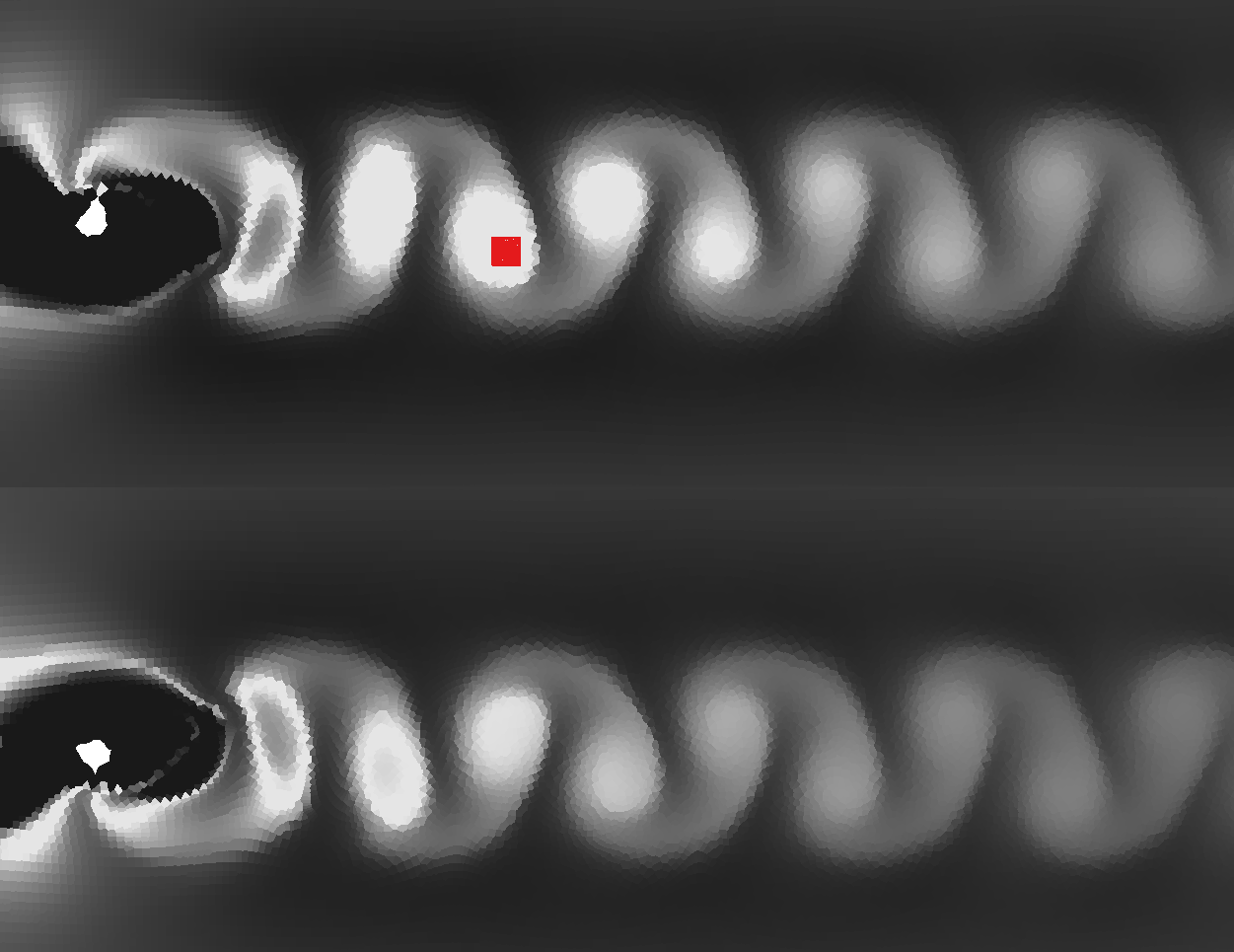}
  }
 
   \caption{Point Similarity of flow over cylinder dataset at $t_0$ (top) and $t_1$ (bottom). (a) similarity based on HKS values in range $[-3.16, -3.65]$. (b) range $[-3.16, 7.00]$ used instead. \label{fig:cylinder_two_timesteps}}
 \end{figure}

We next compare flow over cylinder dataset at different timesteps. The dataset has recurring features that we expect to show up regardless of the initial time stamp.  \autoref{fig:cylinder_two_timesteps} shows the comparison result of the two datasets with different start times $t_0^0$ and $t_0^1$.  The red point is selected near a center of a vortex. In general, vortices at the similar spatial locations are similar in the HKS even when they are at different temporal domain. At small scales, the vortices appear to be more distinct as the HKS at lower scales is more sensitive to local neighborhood. At larger scales, however, the vortices start to blend together and form the vortex street as the HKS gets more influenced by the difference between the street and laminar flow.

\paragraph{Flow over Double Cylinders}

Since the HKS is commensurable, we can compare across flow datasets of similar type. We compare the flow over 2D cylinder dataset with another simulated flow dataset on the same domain of $[-0.5,15.5]\times[-3.5,3.5]$ and same initial velocity. However, we shift the original cylinder with radius $0.125$ upwards to $(4, 1.3)$, and we place another cylinder with radius $0.0625$ below it at $(4, -1.3)$. Therefore, this simulation generates two von K\'{a}rm\'{a}n vortex streets with Reynolds number 160 and 80. We pick the same timestep $t_0 = 55$ with integration duration of $\tau=5.4$ for both datasets. For each pathline, we use $m=150$ uniformly distributed sample timesteps.

\autoref{fig:double_a} shows the result of clustering using k-means with $k=4$. We compare this with clustering on the flow over single cylinder dataset shown in \autoref{fig:cylinder_single_cluster_4}. Flow behind cylinder of the same size shows similar patterns with clustering, while flow behind the smaller cylinder is not captured in the same way.  This difference is explained in more detail in \autoref{fig:double_b} which shows a direct comparison of the HKS curves of points in two vortex streets.

Particularly, in \autoref{fig:double_b} we show HKS curves of five points in the flow over double cylinders dataset. As shown before in the flow over single 2D cylinder dataset (\autoref{fig:example1}), the two points in the vortex street have significantly higher HKS curves than points in laminar region. Moreover, the purple point in the lower vortex street (that has a higher Reynolds number) has a higher HKS because the flow is more tightly concentrated. At small scales the points in laminar regions have similar HKS curves. However, at larger scales the point in the laminar region between two vortex streets start to differ from the other two points because the heat diffusion starts to be influenced by the behavior of the vortex streets.

\begin{figure}[!ht]
  \centering
  \subfloat[$t_0$]{
    \includegraphics[width=0.28\columnwidth]{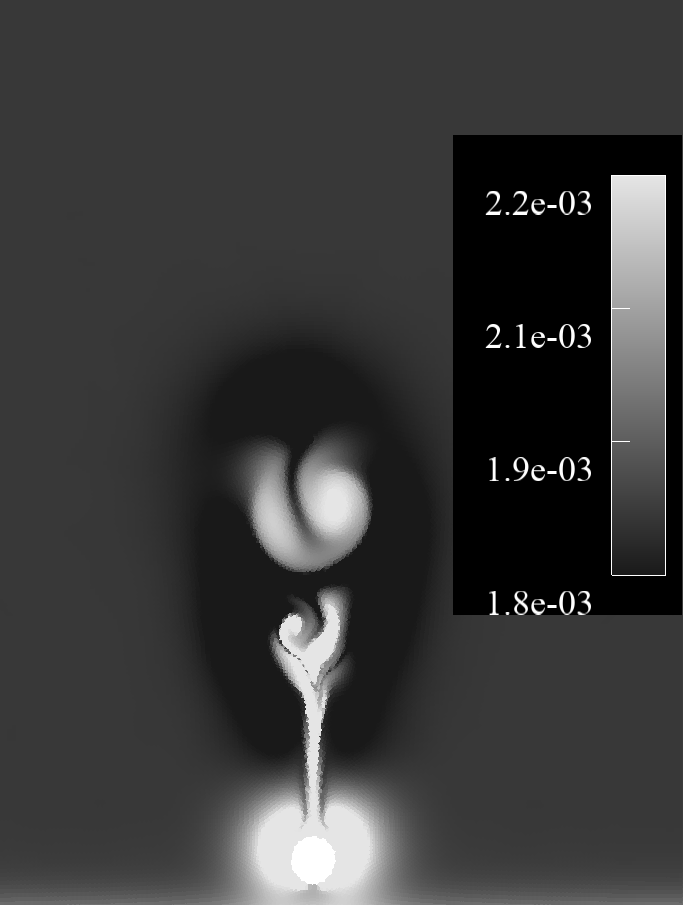}
  }
  \subfloat[$t_1$]{
    \includegraphics[width=0.28\columnwidth]{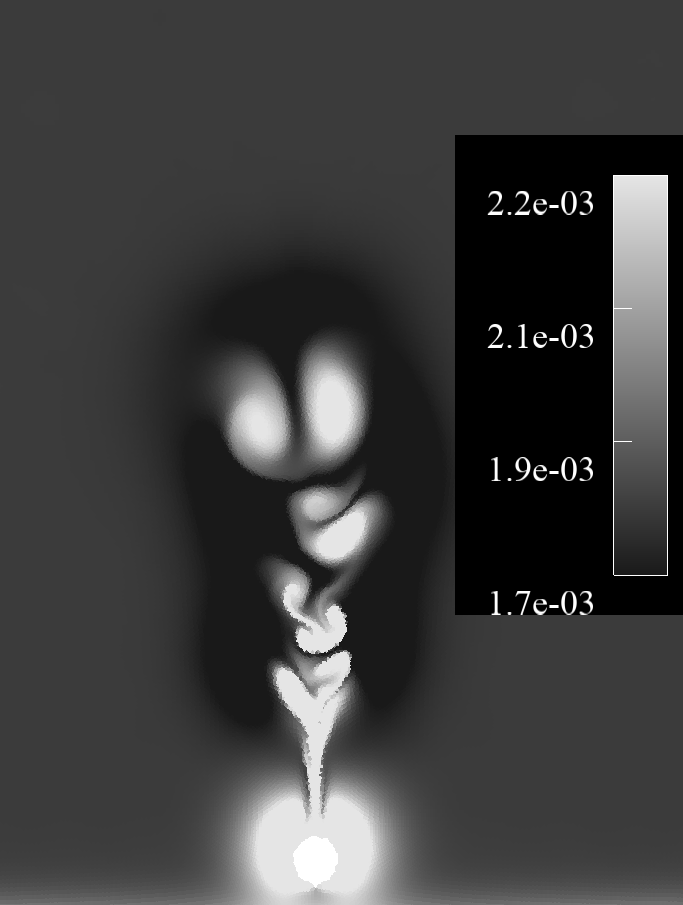}
  }
  \subfloat[$t_2$]{
    \includegraphics[width=0.28\columnwidth]{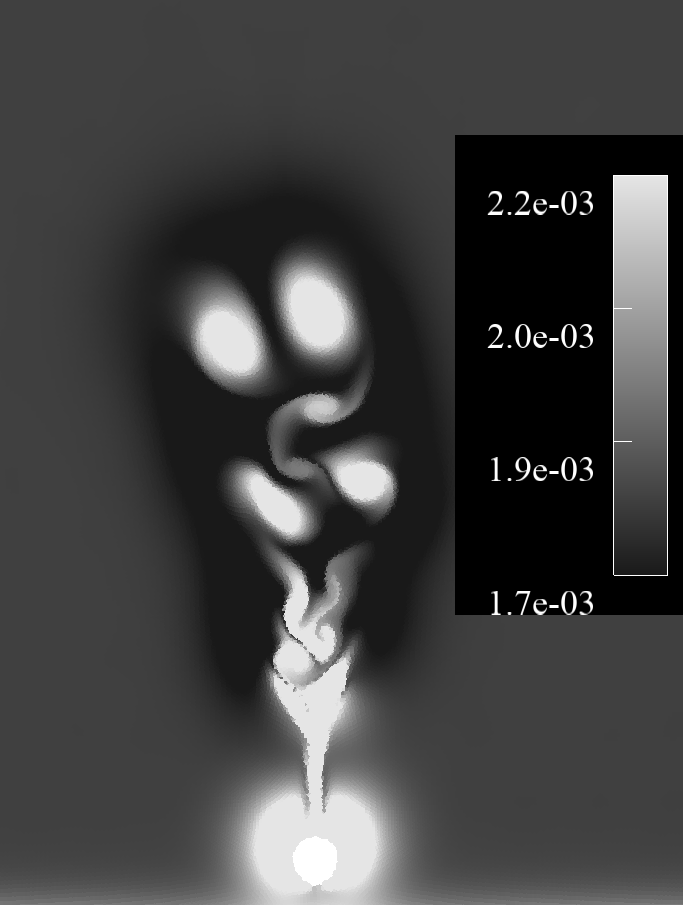}
  }

  \subfloat[$t_0$]{
    \includegraphics[width=0.28\columnwidth]{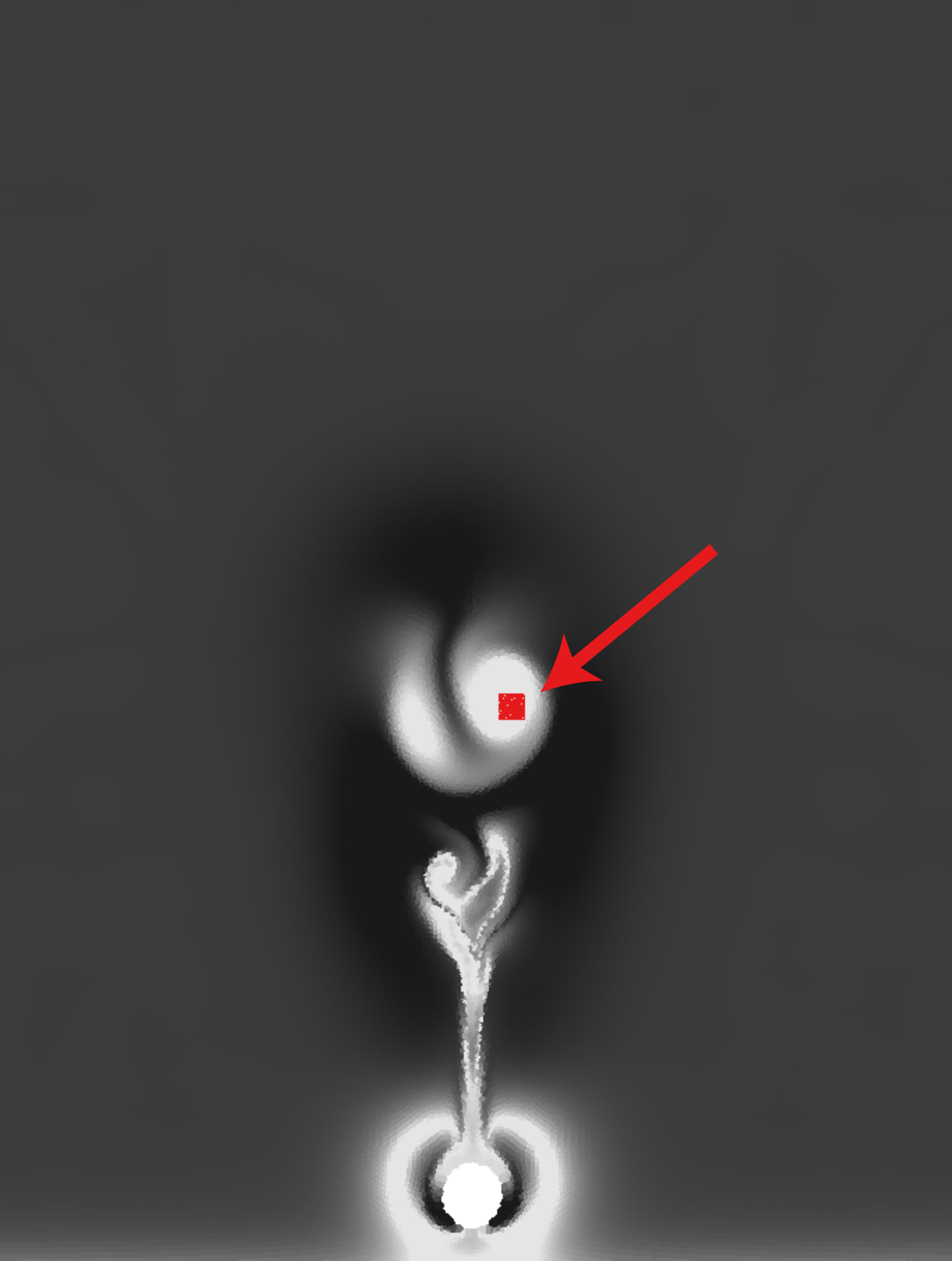}
  }
  \subfloat[$t_1$]{
    \includegraphics[width=0.28\columnwidth]{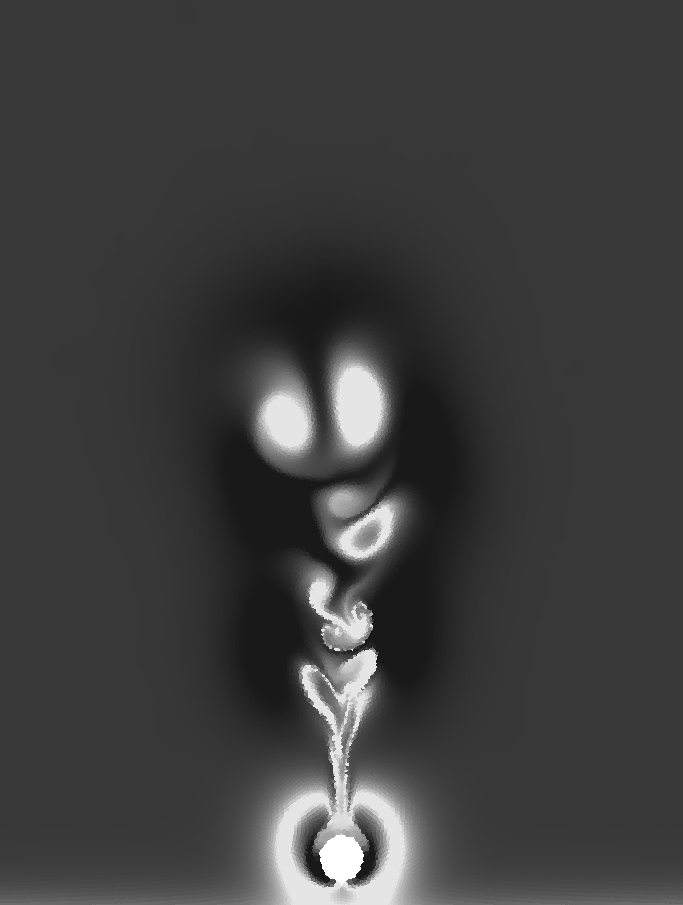}
  }
  \subfloat[$t_2$]{
    \includegraphics[width=0.28\columnwidth]{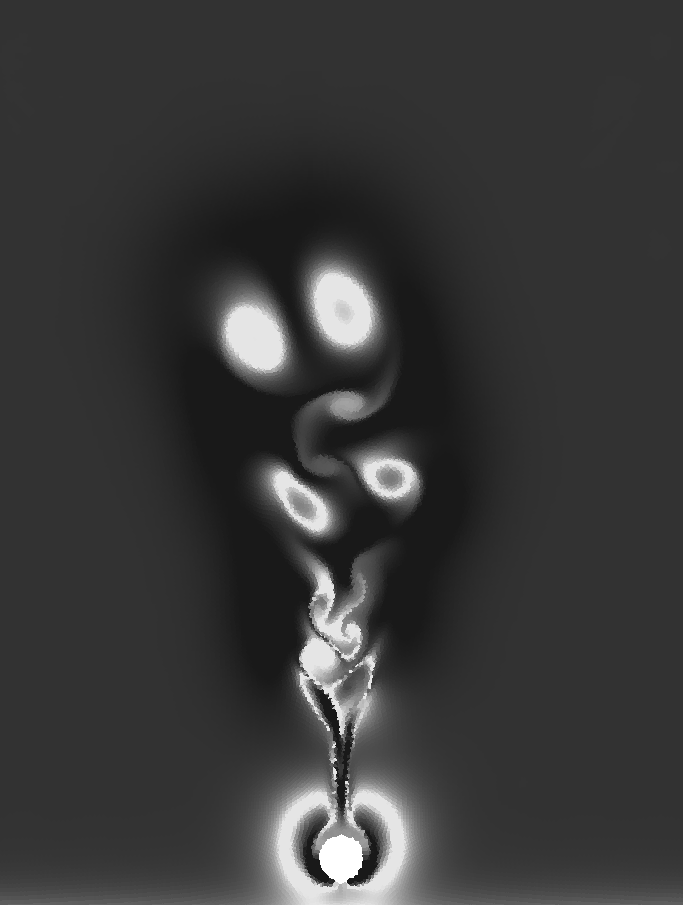}
  }
  
  \caption{\label{fig:heated_evolve} Temporal evolution of the HKS for the heated cylinder with $A_1 = 0$. Top row: Mean HKS with scale range $[0, 0.5]$. Bottom Row: Point Similarity with respect to the red selected point in (d).}
\end{figure}

\paragraph{Convective Flow from a Heated Cylinder}

In this experiment we look at an ensemble of convected flow above a heated cylinder at different angles. This results in a turbulent plume and the formation of vortices nonuniformly distributed throughout the domain. The resulting Boussinesq flow is simulated using Gerris on a domain of $[-1.5,1.5]^2$ with a cylinder of radius 0.16 centered at $(0,-0.15)$. Each dataset in the ensemble varies the direction where the heat diffuses initially. Specifically, we set diffusion source vector $(u,v)$ so that $u = \cos(\frac{A_i}{180}\pi)$ and $v = \sin(\frac{A_i}{180}\pi)$, where $A_i$ is the parameter for each dataset in the ensemble. We choose $A_0 = -3, A_1 = 0, A_2 = 3$ for the datasets in this experiment.  For each possible angle parameter, we consider pathlines that start at $t_0^0=4.73$, $t_0^1=5.77$, $t_0^2=6.86$  with integration duration of $\tau=0.85$. For each pathline, we use $m=50$ uniformly distributed sample timesteps.


For a single dataset in the ensemble, we analyze different stages as the plume above cylinder evolves. \autoref{fig:heated_evolve} shows a comparison of different timesteps of heated cylinder with $A_1 = 0$. In the top row, we see similar patterns of the HKS with flow over cylinder. i.e.~swirling vortices have high HKS values while laminar flow have low values.  These vortices are consistently similar as measured through the mean HKS even though their shape distorts over time.  Points near the turbulent region form a distinct region with even lower values than laminar region. 

In the bottom row of \autoref{fig:heated_evolve}, we highlight points that are similar with the point selected in a vortex at an early timestep. We see the vortices in later timesteps are highlighted in Point Similarity view.  Just as with the flow over cylinder dataset, the HKS encodes intrinsic properties of the flow like the size of the vortices, therefore our viewer highlights, at different timesteps, vortices that are similar in size. As the same vortex (e.g.~the one with the red selected point) evolves over time and eventually dissipates, its HKS will gradually decrease and converge to laminar flow.

\begin{figure}[!t]
  \centering
  \subfloat[$A_0$]{
    \includegraphics[width=0.28\columnwidth]{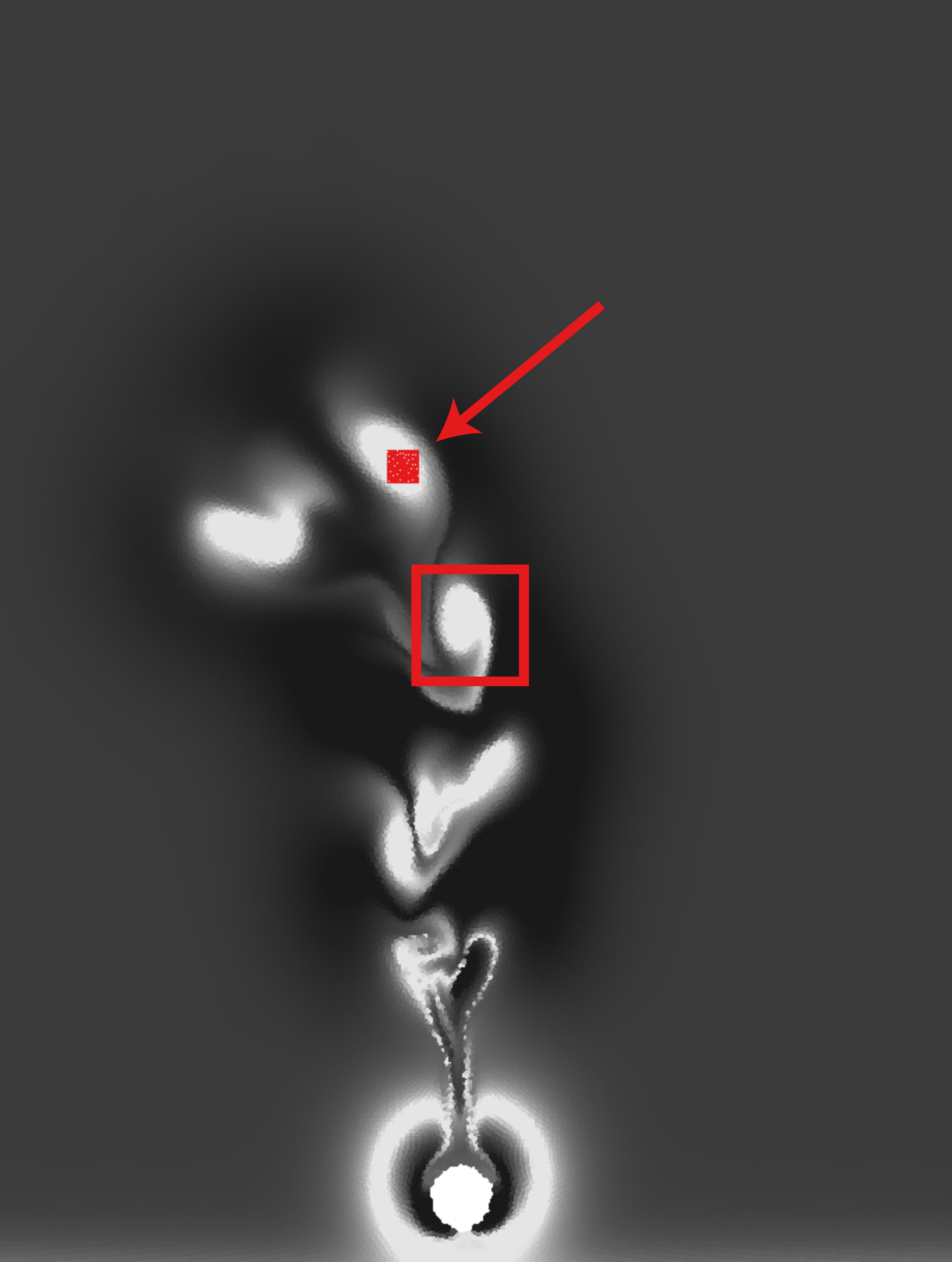}
  }
  \subfloat[$A_1$]{
    \includegraphics[width=0.28\columnwidth]{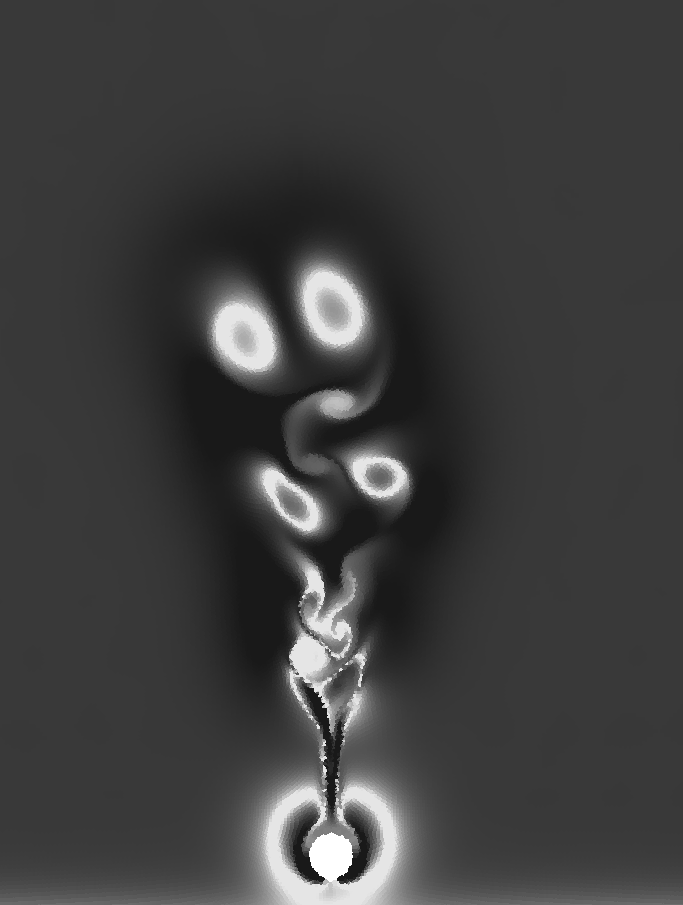}
  }
  \subfloat[$A_2$]{
    \includegraphics[width=0.28\columnwidth]{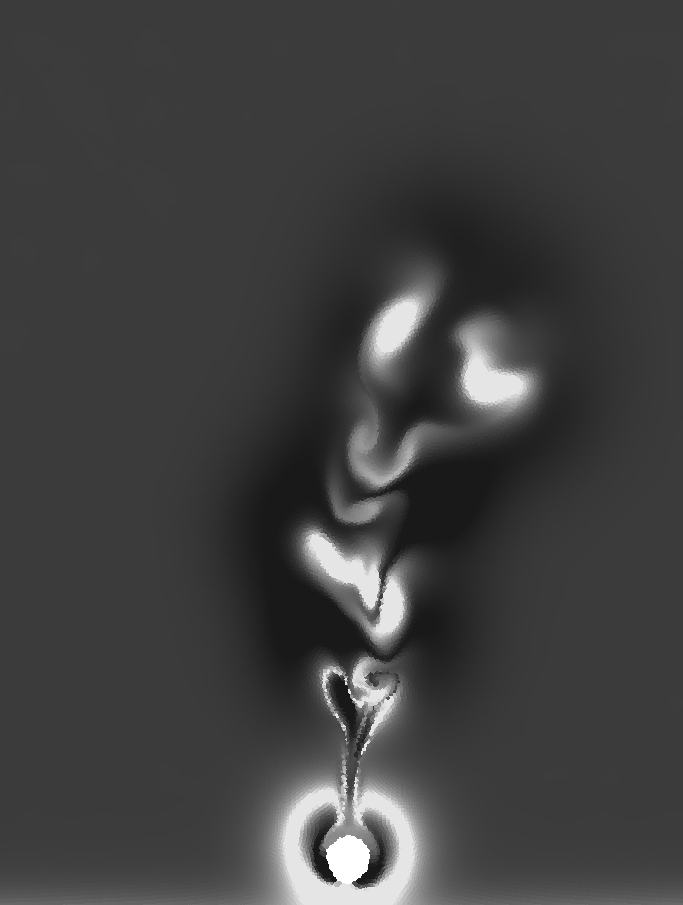}
  }

  \subfloat[$A_0$]{
    \includegraphics[width=0.28\columnwidth]{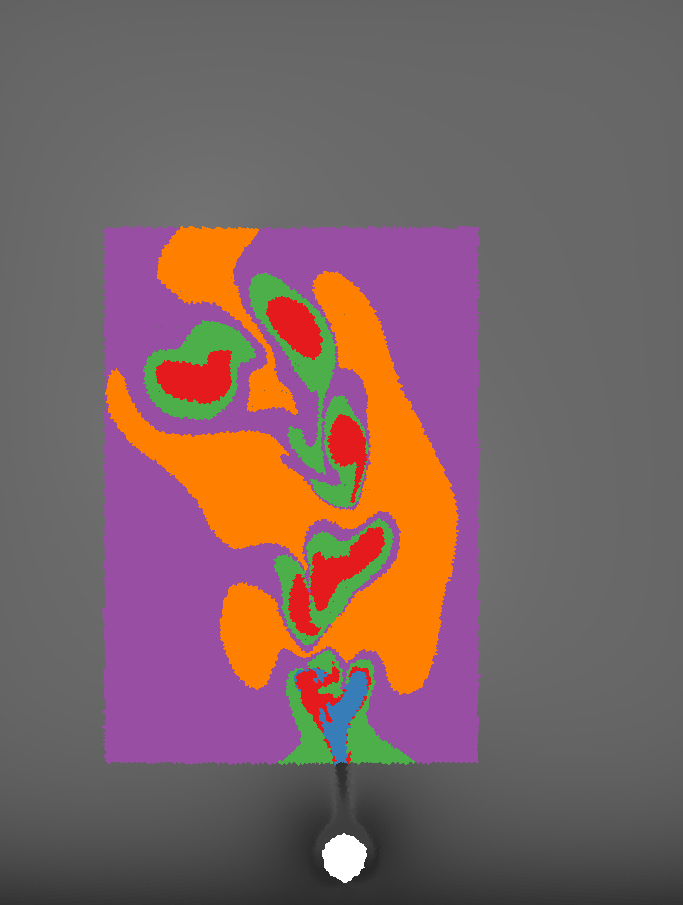}
  }
  \subfloat[$A_1$]{
    \includegraphics[width=0.28\columnwidth]{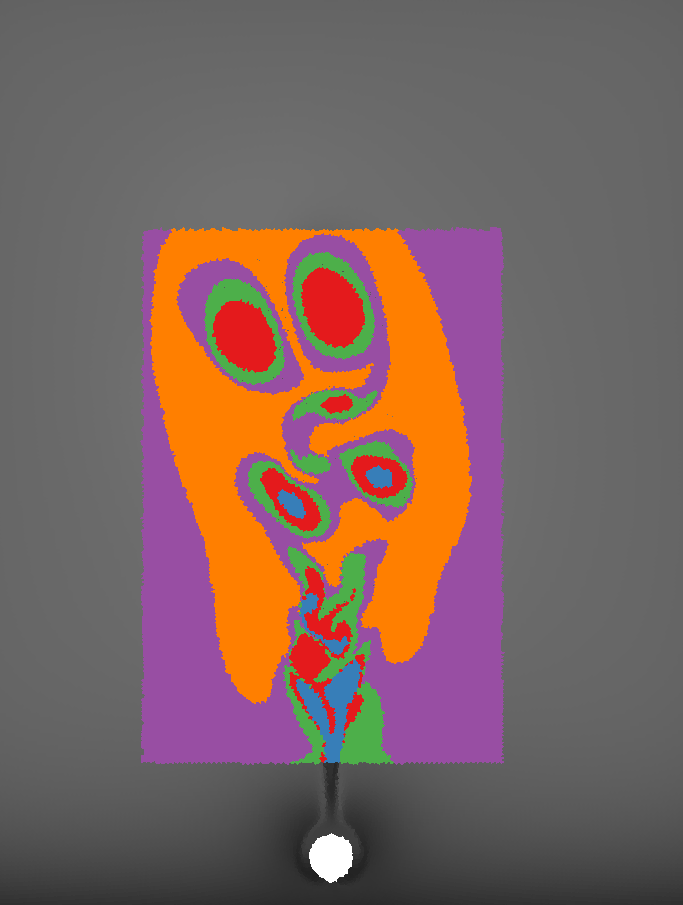}
  }
  \subfloat[$A_2$]{
    \includegraphics[width=0.28\columnwidth]{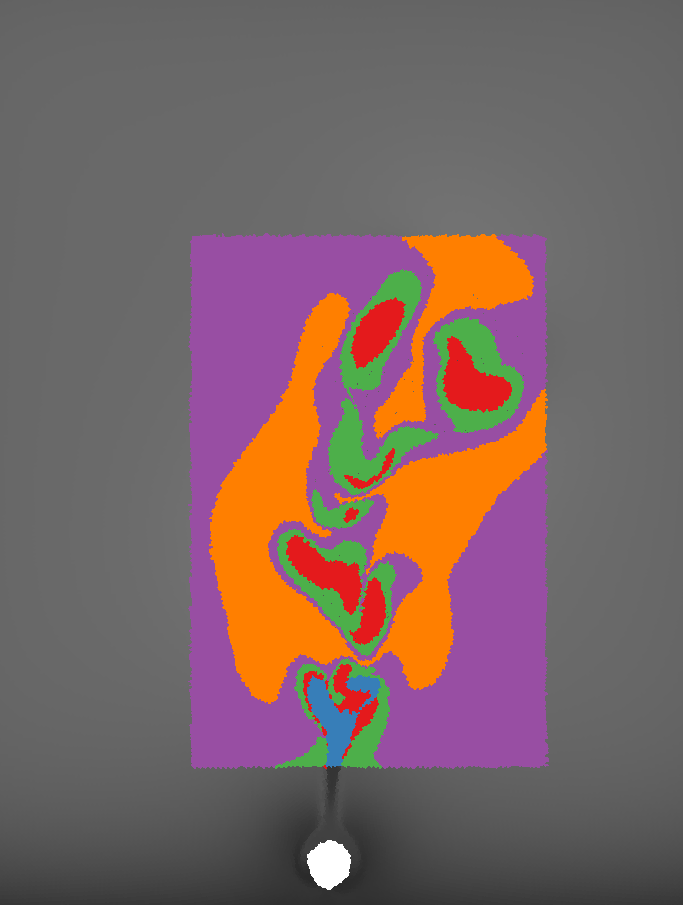}
  }
  
  \caption{\label{fig:heated_compare} Comparison of an ensemble where heat convects in different directions. Top row: Points Similarity with respect to the red point in $A_0$ Bottom row: HKS Clustering with 4 clusters.}
\end{figure}

We next compare across ensemble datasets. \autoref{fig:heated_compare} shows the results of comparing three datasets from the ensemble. $A_0$ and $A_2$ are mostly symmetric, and this fact is highlighted with our Point Similarity view, as shown in the top row of \autoref{fig:heated_compare}. $A_0$ does have a vortex (in the red box) that does not have a symmetric correspondence in $A_2$, but it has two very similar ones in $A_1$. These relationships are all captured in our viewer. The bottom row of \autoref{fig:heated_compare} shows another example of our clustering view for $k=4$, which separates the vortices and the laminar region and shows correspondence of regions for different datasets in the ensemble, even though the shapes of these vortices are different.

\paragraph{Robustness to Spatial Distribution of Pathlines}

\begin{figure}[!ht]
  \centering
  \subfloat[]{
    \includegraphics[width=0.22\columnwidth]{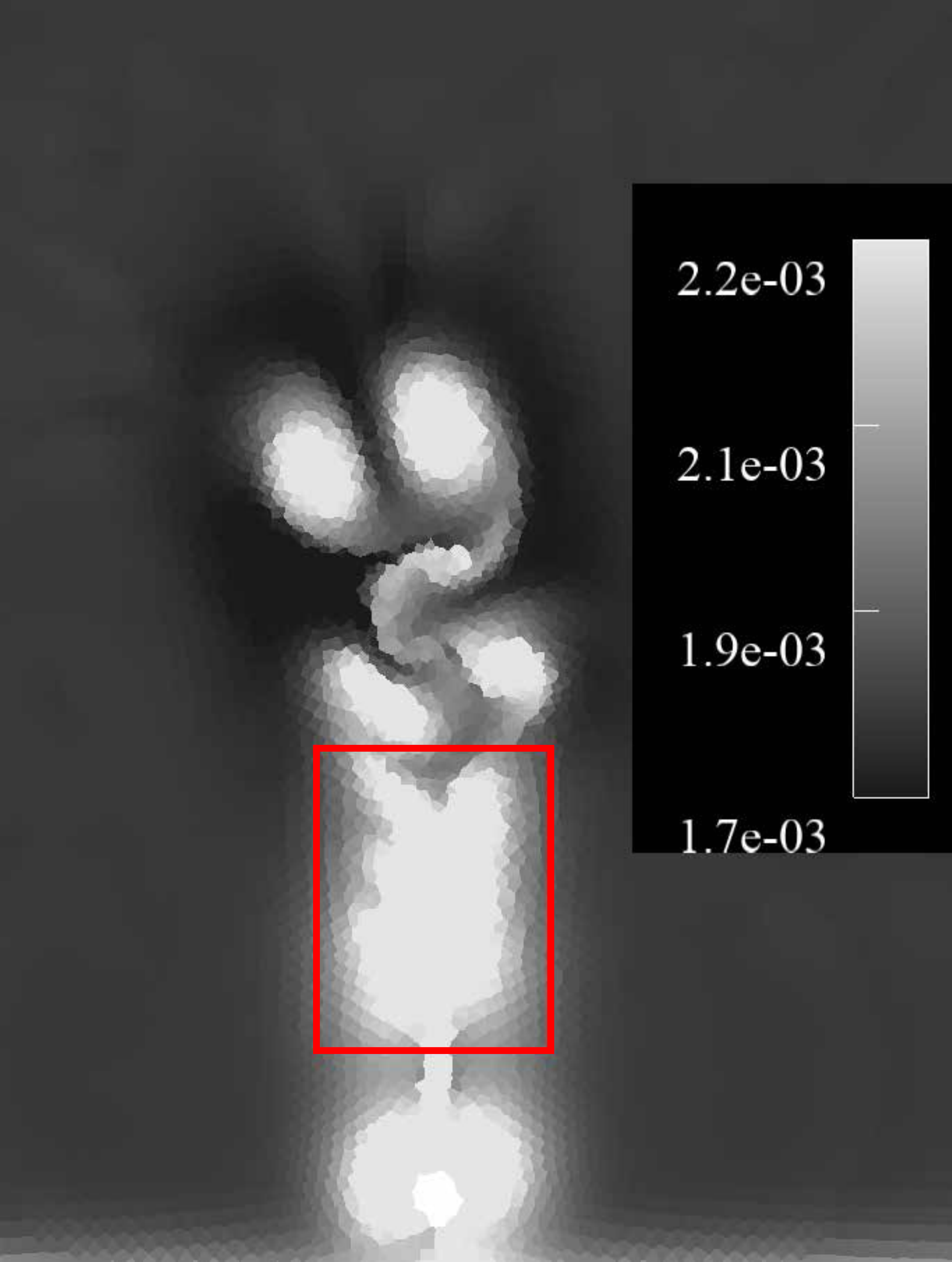}
  }
  \subfloat[]{
    \includegraphics[width=0.22\columnwidth]{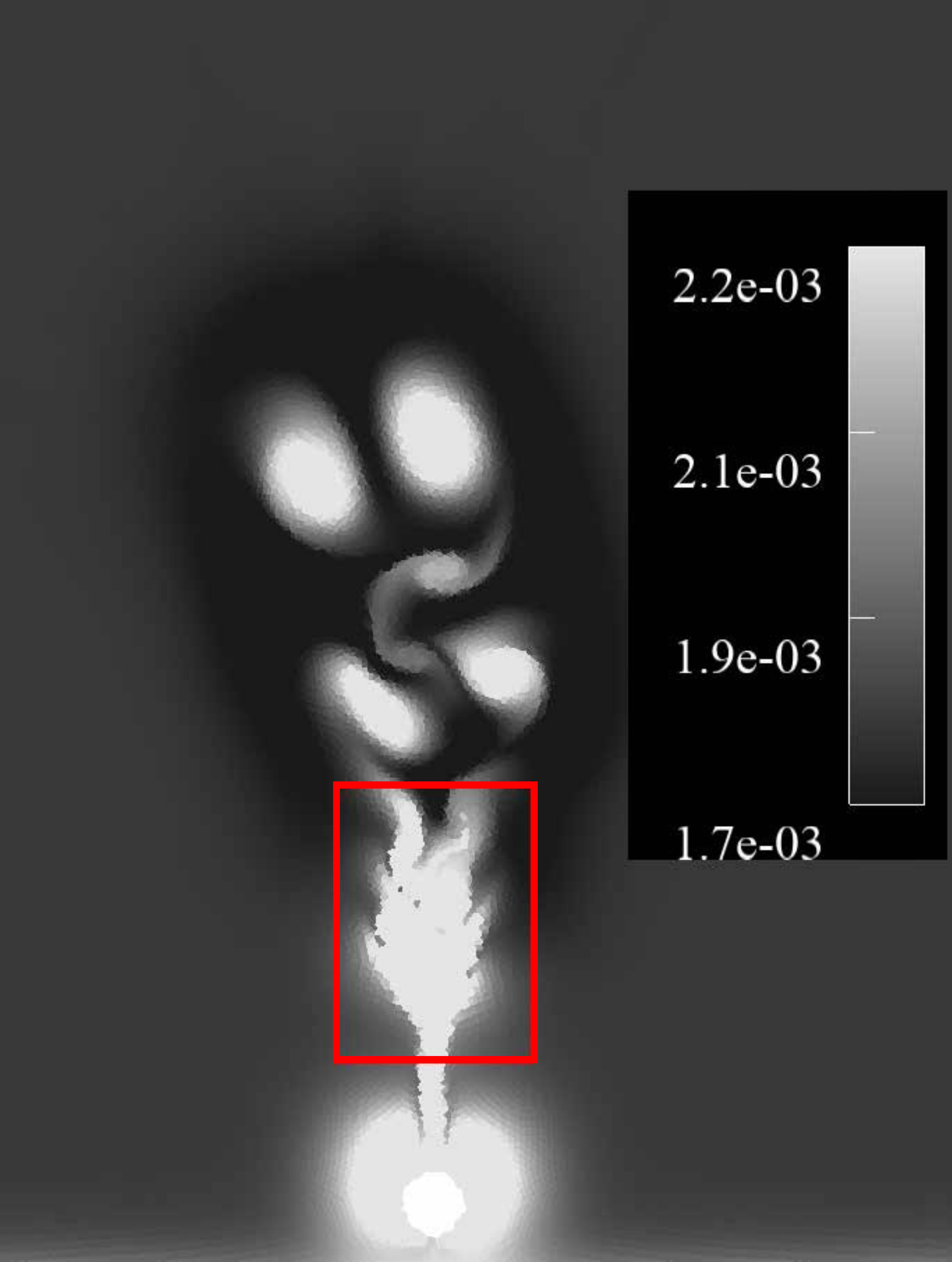}
  }
  \subfloat[]{
    \includegraphics[width=0.22\columnwidth]{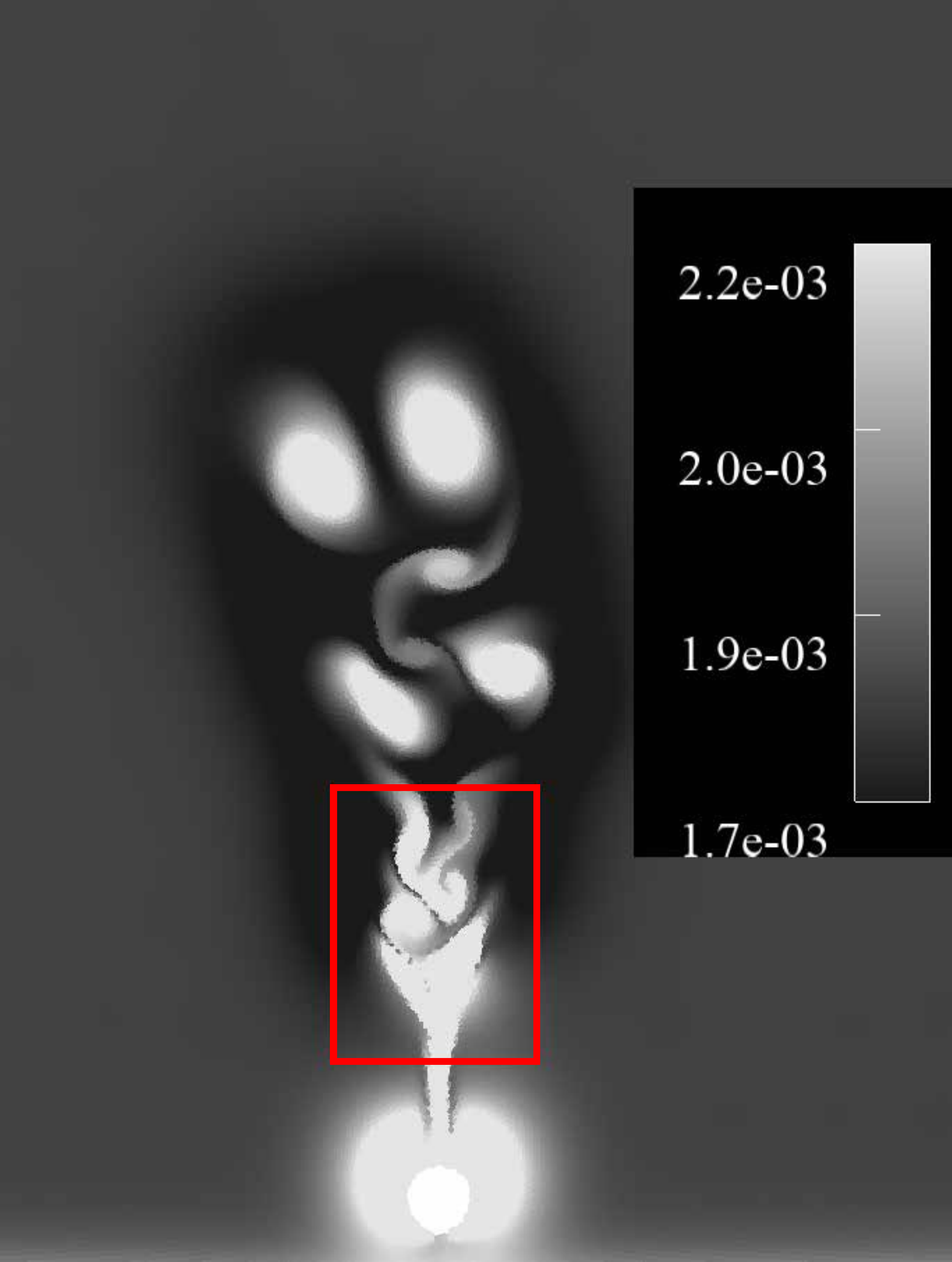}
  }
  \subfloat[]{
    \includegraphics[width=0.22\columnwidth]{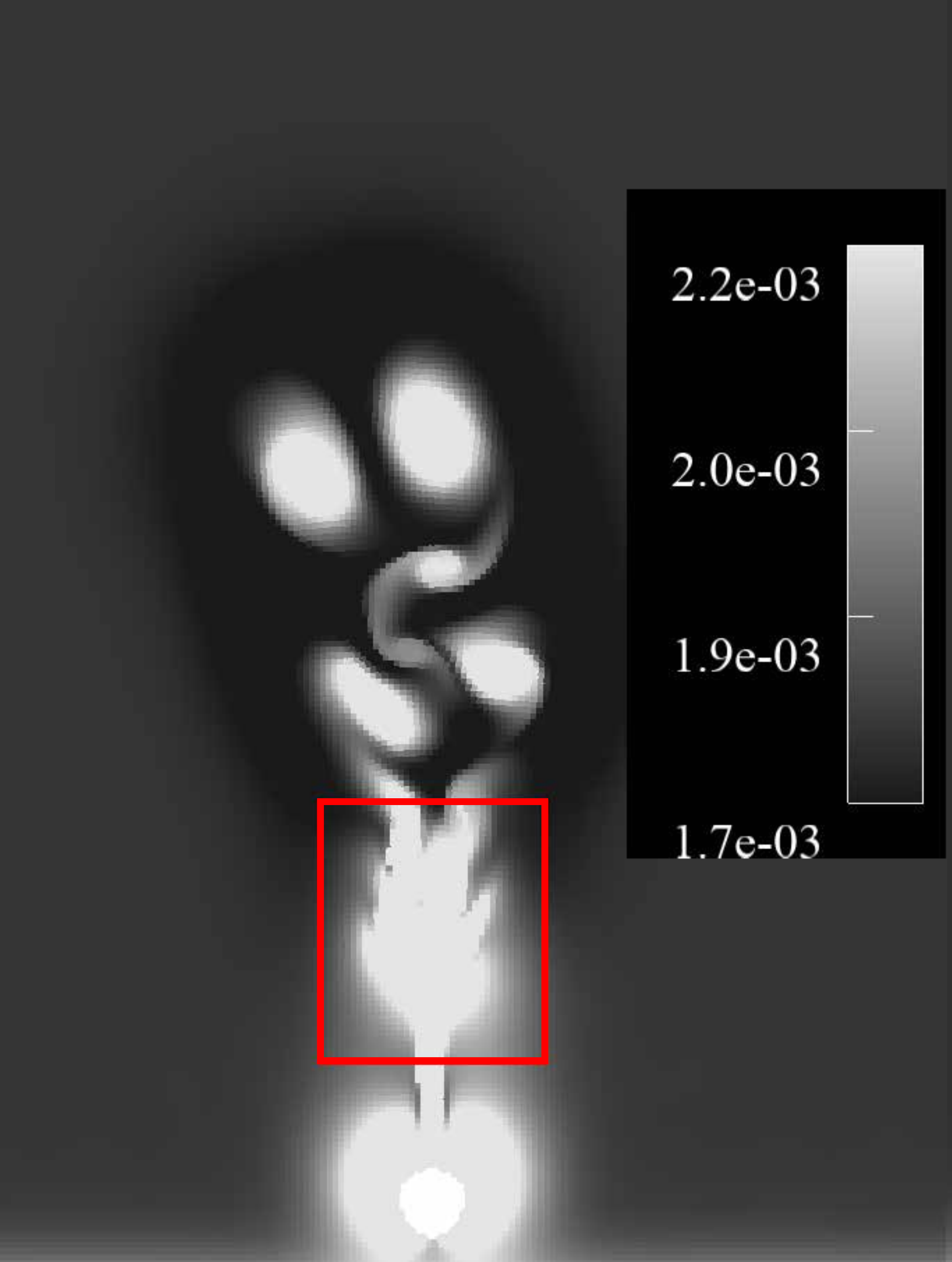}
  }

  \subfloat[]{
    \includegraphics[width=0.22\columnwidth]{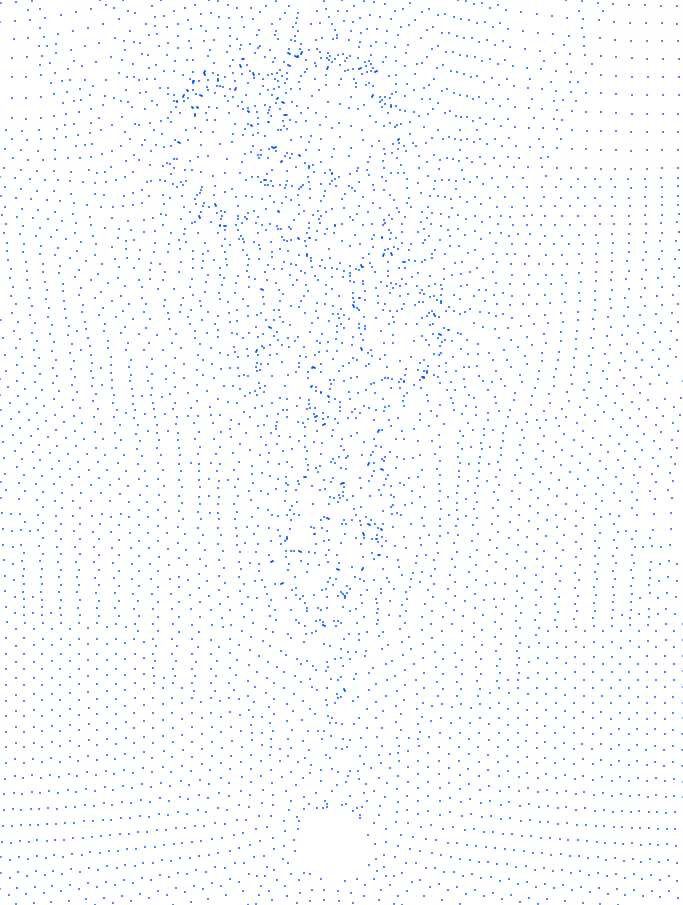}
  }
  \subfloat[]{
    \includegraphics[width=0.22\columnwidth]{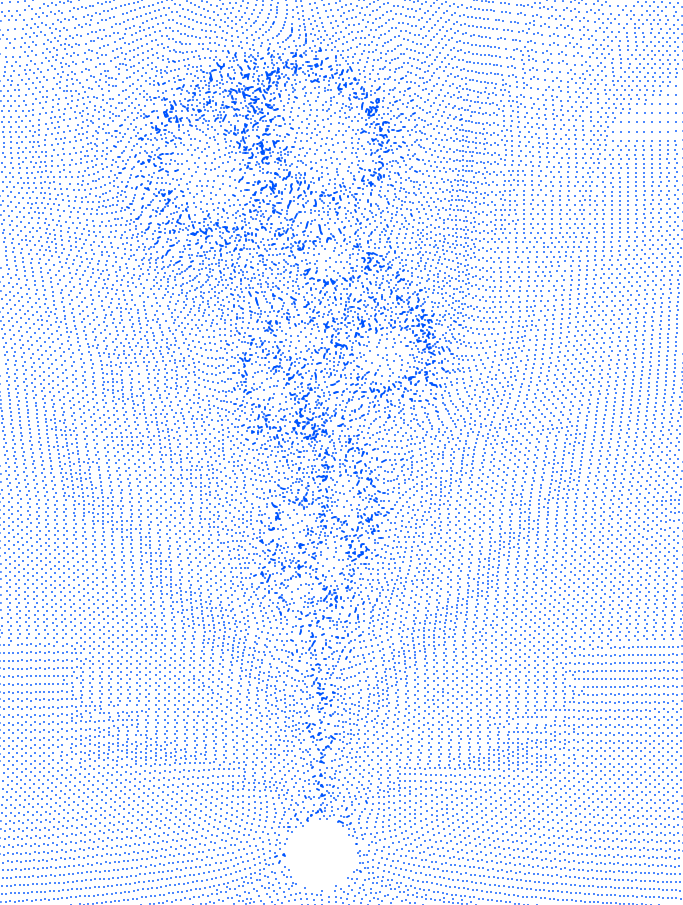}
  }
  \subfloat[]{
    \includegraphics[width=0.22\columnwidth]{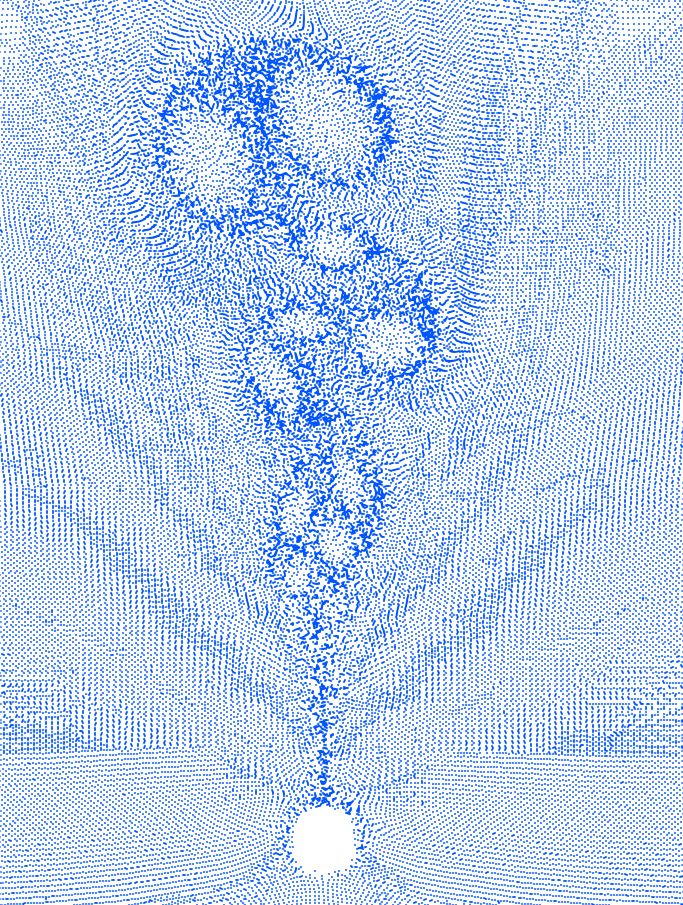}
  }
  \subfloat[]{
    \includegraphics[width=0.22\columnwidth]{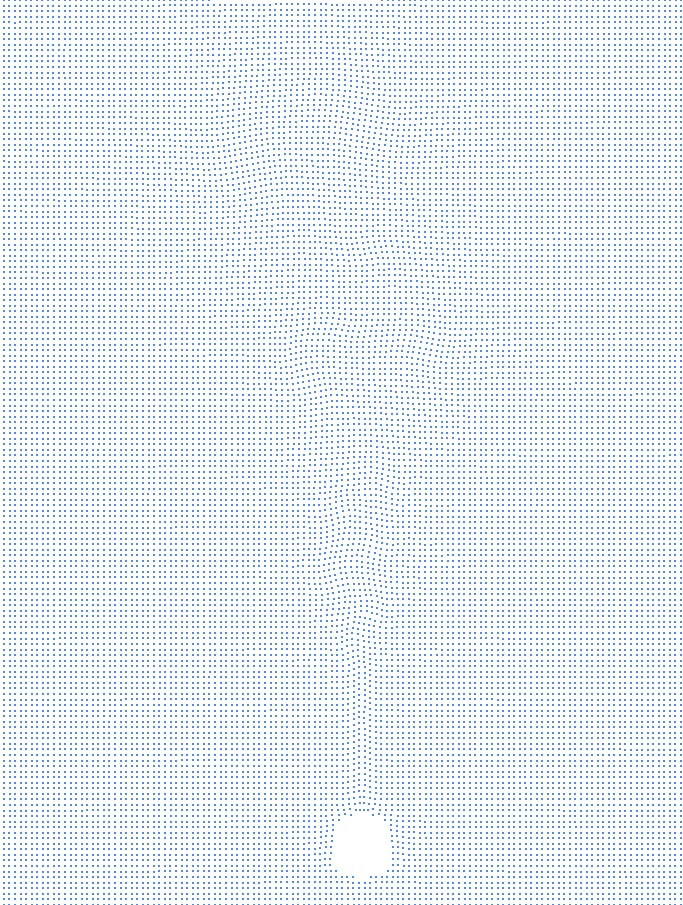}
  }
  \caption{\label{fig:heated_density} A comparison where both number and spatial distribution of pathlines vary. Top: Mean HKS of the $A_0$ heated cylinder dataset with different spatial distribution of pathlines. Bottom: Starting positions of the pathlines in the above sample. Number of pathlines: (a) 16572 (b) 74205 (c) 175518 (d) 123493.}
\end{figure}

Finally, we show that our technique is stable for different spatial distribution of pathlines, including varying density and uniform or nonuniform distributions. \autoref{fig:heated_density} shows a comparison of the $A_0$ heated cylinder dataset with different number of pathlines as well as uniform versus nonuniform pathline distributions. In most laminar region and vortices, the HKS is nearly identical with different distribution of pathlines. However, the HKS for points in the red box region in \autoref{fig:heated_density} varies for different distributions of pathlines. This is because that in such highly turbulent area, lower number of pathlines will be insufficient to fully capture the local feature of the neighborhood.  Otherwise, the HKS is stable for varying number of pathlines as well as uniform or nonuniform spatial distributions.

\section{Discussion}

In this work, we showed that the heat kernel signature can be used for visual analysis of unsteady flow.  Since the HKS captures intrinsic geometry in a multi-scale way, it allows for a detailed comparison of pathlines that can provide correspondence across different timesteps and even different simulations. The HKS can be viewed as a high-dimensional feature, but each dimension of this feature can provide various insights to the data.  

Computationally, there are some limitations to the work in that our procedure involves estimating and eigendecomposing the Laplace-Beltrami operator, the size of which relates to the number of samples that are used and the dimension by which affinity is estimated. 
While our experiments support that the pathline manifold $P$ appears to have a fixed dimension, additional experimentation is needed on more datasets to confirm both what the dimension is (for high-dimensional flows, it likely is not two-dimensional) and whether or not there exist datasets where the overwhelming majority of pathlines do not have the same dimension.
Further, while there is a trade off with sparsification of this matrix and accuracy of the heat kernel, we found that for datasets of size hundreds of thousands of particles, it was still reasonable to compute it as a preprocess. We leave for future work improving the computation process including parameter tuning and a more comprehensive study on the dimension analysis of the pathline manifold. 

A more pressing challenge is that the pathline manifold may have high-dimensional boundaries that influence the results we see.  Like past work, our goal was not to contribute a new method for manifold learning, but rather to rely on current approaches and understand their strengths and weaknesses.  Boundaries have a direct impact on the trade off between pathline density (viewed through $\sigma$) and diffusion scale $s$.  Particularly, near boundaries we see some distortion, especially at large scales.  Related, it becomes more difficult to accurately estimate volumes near boundaries.  Nevertheless, we found our estimates of the HKS to behave correctly for pathlines sufficiently far from the boundary, suggesting that appropriately padding the data may help. 


The features we extracted characterize the geometry of the pathline manifold through the heat kernel.  The next steps of this work are to understand the connections between these shape analysis features and true physical properties of the underlying flow.  While some authors have already made connections~\cite{banisch2017understanding,froyland2016dynamic,karrasch2016geometric}, additional work is necessary to fully explore whether geometric features can be mapped to specific physical properties of the flow.

Finally, while in this work we did not experiment in 3D, the methods we use should be applicable.  A major challenge is estimating volumes for higher dimensional manifolds, since the local dimensionality might increase, but most of the HKS computational pipeline can translate.  Still, even with the HKS values in hand for a 3D pathline dataset, interpreting them could be quite challenging.  We leave for future work the significant modifications to our visual interface that this would require, but note that filtering on the HKS curves themselves offers a promising direction to explore.

\acknowledgments{
  This material is based upon work supported by the U.S. Department of Energy, Office of Science, Office of Advanced Scientific Computing Research, under Award Number(s) DE-SC-0019039.}

\bibliographystyle{abbrv-doi}

\bibliography{pvis20_hks,josh}
\end{document}